\documentclass[aps,prl,twocolumn,floatfix,superscriptaddress]{revtex4}

\usepackage{amssymb,amsmath,amstext}                %%   American Physical Society math etc extensions
\usepackage{graphicx}
\usepackage{epstopdf}                                               %%   help with eps -> pdf
\usepackage{color}                                                     %%   change color
\usepackage{bm}                                                        %%   bold math
\usepackage{appendix}

\usepackage[utf8]{inputenc}
\usepackage{lipsum}

\usepackage{ulem}
\usepackage{latexsym}
\usepackage[colorlinks=true,citecolor=blue,linkcolor=magenta]{hyperref}
\usepackage{soul}
\usepackage{amsmath}
\usepackage{float}

\def\be{\begin{equation}}
\def\ee{\end{equation}}
\def\bea{\begin{eqnarray}}
\def\eea{\end{eqnarray}}
\def\ra{\rangle}
\def\la{\langle}
\def\bi{\begin{itemize}}
\def\ei{\end{itemize}}

\definecolor{dgreen} {RGB}{78,138,21}

\definecolor{xuedong} {RGB}{0,128,0}

\definecolor{giergiel} {RGB}{0,162,108}

\definecolor{purple} {RGB}{128,0,160}

\begin{document}
\title{Discrete Time Crystals with Absolute Stability}
\author{Krzysztof Giergiel}
\affiliation{Instytut Fizyki Teoretycznej, Uniwersytet Jagiello\'{n}ski, ulica
Profesora Stanislawa Lojasiewicza 11, PL-30-348 Krak{ó}w, Poland}
\affiliation{Optical Sciences Centre, Swinburne University of Technology, Melbourne
3122, Australia}
\author{Jia Wang}
\affiliation{Centre for Quantum Technology Theory, Swinburne University of Technology,
Melbourne 3122, Australia}
\author{Bryan J. Dalton}
\affiliation{Centre for Quantum Technology Theory, Swinburne University of Technology,
Melbourne 3122, Australia}
\author{Peter Hannaford}
\affiliation{Optical Sciences Centre, Swinburne University of Technology, Melbourne
3122, Australia}
\author{Krzysztof Sacha}
\affiliation{Instytut Fizyki Teoretycznej, Uniwersytet Jagiello\'{n}ski, ulica
Profesora Stanislawa Lojasiewicza 11, PL-30-348 Krak{ó}w, Poland}
\begin{abstract}
We show that interacting bosons on a ring which are driven periodically by a rotating potential can support discrete time crystals whose absolute stability can be proven. The absolute stability is demonstrated by an exact mapping of discrete time crystal states to low-lying eigenstates of a time-independent model that reveals spontaneous breaking of space translation symmetry. The mapping ensures that there are no residual time-dependent terms that could lead to heating of the system and destruction of discrete time crystals. We also analyze periodically kicked bosons where the mapping is approximate only and cannot guarantee the absolute stability of discrete time crystals. Besides illustrating potential sources of instability, the kicked bosons model demonstrates a rich field for investigating the interplay between different time and space symmetry breaking, as well as the stability of time crystal behavior in contact with a thermal reservoir.
\end{abstract}
\date{\today}

\maketitle

Ordinary space crystals correspond to periodic distributions of particles in space which form despite the fact that the Hamiltonian is invariant under a translation of all the particles by an arbitrary vector. This is the phenomenon of spontaneous breaking of the continuous space translation symmetry into a discrete space translation symmetry in a time-independent many-body system. In 2012, research on time crystals was initiated where spontaneous breaking of time translation symmetry is responsible for the formation of a crystalline structure~\cite{WilczekPRL2012}. Periodically driven many-body closed systems can form discrete time crystals as a non-equilibrium quantum state in which the stable system response occurs at integer multiples of the drive period~\cite{Sacha2015PRA,KhemaniPRL2016,ElsePRL2016,YaoPRL2017}. The discrete time translation symmetry of a time-periodic Hamiltonian is spontaneously broken and new periodic motion emerges --- a novel crystalline structure appears in time~\cite{Sacha2015PRA}. 

The stability of discrete time crystals is much less obvious than for space crystals because the quasi-energy spectrum of the Floquet Hamiltonian has no lower bound~	\cite{SachaTC2020}. The key question arises whether discrete time crystals are really stable or they gradually absorb energy from the drive and eventually heat up to a structureless infinite temperature state as expected for a generic periodically driven many-body system~\cite{DAlessioPRX2014,LazaridesPRE2014,PonteAPA2015}. 

Systems which do not thermalize are typically integrable systems but integrability is fragile and usually requires fine-tuning of system parameters \cite{RigolPRL2007,SantosPRE2010a,MarcosPRA2010,SantosPRE2010b,CassidyPRL2011,CauxPRL2013}. Time-independent many-body systems in the presence of disorder are believed to reveal many-body localization (MBL) and emergent integrability \cite{Nandkishore2015,Altman2015,DAlessio2016,Mierzejewski2018}. Therefore, periodically driven MBL systems are potential candidates for realization of discrete time crystals \cite{ElsePRL2016,KhemaniPRL2016,ElsePRX2017}. However, the stability of discrete time crystals in driven MBL systems  \cite{PontePRL2015,LazaridesPRL2015,AbaninAP2016,Sierant2023} is far from being obvious \cite{ZhangNature2017,ChoiNature2017,
Pal2018,Rovny2018,Smits2018,Mi2022,
Randall2021,Frey2022} --- even in the time-independent case, the MBL is still debated \cite{Imbrie2016,Sels2021}. Also, the stability of the first proposed discrete time crystal in ultra-cold bosonic atoms bouncing resonantly on an oscillating atom mirror \cite{Sacha2015PRA} has not been proven. While previous studies have demonstrated the stability of discrete time crystals for evolution times longer than those of realistic experiments \cite{WangNJP2021,Kuros2020,WangPRA2021}, their validity in the context of an infinite number of bosons and infinite evolution time remains unproven. 

Here, we show that resonantly driven bosons by a rotating potential on a ring in the Lieb-Liniger (LL) model can reveal discrete time crystals whose stability can be proven. Eigenstates of the Floquet Hamiltonian that break the discrete time translation symmetry can be mapped to low energy eigenstates of a time-independent Hamiltonian in the rotating frame. The latter eigenstates are stable and reveal spontaneous breaking of \emph{space} translation symmetry in the rotating frame, which corresponds to spontaneous breaking of the discrete \emph{time} translation symmetry of the original model. This model concretely provides an example of the absence of quantum thermalization and the existence of stable time crystals in a closed quantum many-body system without the need for either disorder or integrability. Furthermore, an experimental realisation of this model should be possible, since ultra-cold atoms on a ring have been created and applying a periodic drive potential should be straightforward \cite{SM,Gupta2005,bell2016,moulder2012,kumar2016}.

Let us consider $N$ bosons with contact interactions on a ring with circumference $2\pi$ which are periodically driven by a rotating potential
\bea
H&=&H_{\rm LL}+\sum_{i=1}^N V(2x_i-\omega t), %\lambda\sum_{i=1}^N \cos(2x_i-\omega t),
\label{ll1} \\
H_{\rm LL}&=&\sum_{i=1}^N\frac{p_i^2}{2}+g_0\sum_{i<j}^N\delta(x_i-x_j),
\label{ll1a}
\eea
where we use $R$ and $\hbar^2/mR^2$ for the length and energy units, respectively, where $R$ is the ring radius and $m$ the mass of the bosons and $g_0$ stands for the strength of the contact interactions \cite{footnote1}. We assume that the potential in (\ref{ll1}) has a double-well structure which remains unchanged if $x_i\rightarrow x_i+\pi$. Apart from such a discrete space translation symmetry, the Hamiltonian possesses also a discrete time translation symmetry, i.e., $t\rightarrow t+T$ (where $T=2\pi/\omega$).  We will see that for sufficiently strong interactions, these symmetries can spontaneously be broken and the system starts evolving with a period of $2T$ forming a discrete time crystal.

Let us investigate the system in the moving frame of the rotating potential, where we first perform a time-dependent unitary transformation $U_t = \exp(i\sum_jp_j\omega t/2)$, leading to a shift in the positions, $x_i \rightarrow x_i+\omega t / 2$, and next a second time-independent unitary transformation $U_p=\exp \left(-i \sum_j x_j \omega / 2\right)$, leading to a shift in the momenta, $p_i\rightarrow p_i+\omega/2$ \cite{footnote2}. Under these transformations, we end up with the following exact time-independent Hamiltonian
\be
\tilde H=\sum_{i=1}^N\left[\frac{p_i^2}{2}+V(2x_i)%\lambda \cos(2x_i)
\right]+g_0\sum_{i<j}^N\delta(x_i-x_j),
\label{ll2}
\ee
where a constant term has been omitted. The state vector in the moving frame is related
to that in the laboratory frame via $\tilde\psi=
U_pU_t \psi$ \cite{SM}.

Suppose that the discrete space translation symmetry of the Hamiltonian (\ref{ll2}), i.e., the invariance under the shift $x_i\rightarrow x_i+\pi$, is spontaneously broken in the ground state and low-energy eigenstates and only the symmetry related to the periodic boundary conditions on a ring remains. The corresponding time-independent single-particle probability densities in the moving frame, $\tilde P(x)=\int dx_2\dots dx_N|\tilde\psi(x,x_2,\dots,.x_N)|^2$, fulfill the periodic boundary conditions, $\tilde P(x+2\pi)=\tilde P(x)$, but $\tilde P(x+\pi)\ne \tilde P(x)$. When we return to the original laboratory frame by means of the inverse $U_p$ and $U_t$ transformations, the lab probability densities read $P(x,t)=\tilde P(x-\omega t/2)$ \cite{SM} and thus, due to the spontaneous breaking of the space translation symmetry of the Hamiltonian (\ref{ll2}), they now also reveal spontaneous breaking of the discrete time translation symmetry of the Hamiltonian (\ref{ll1}), since $P(x,t)$ is periodic with the period $2T$ but not with the period $T$. Hence, the system spontaneously starts evolving with a period which is an integer multiple of the period dictated by the drive and forms a discrete time crystal. 

Spontaneous breaking of the space translation symmetry of the Hamiltonian (\ref{ll2}) occurs in the thermodynamic limit (i.e., for $N\rightarrow\infty$, $g_0\rightarrow 0$ but $g_0N=\rm const$ and the circumference of the ring is always equal to $2\pi$). In this limit the ground state of (\ref{ll2}) is a Bose-Einstein condensate (BEC) where all bosons occupy the same single particle wavefunction $\phi_0(x)$ which is the solution of the mean-field Gross-Piteavskii equation \cite{Pethick2002}. If $V$ is a symmetric double-well potential, then for sufficiently strong attractive interactions (i.e., for sufficiently negative $g_0N$), there are two degenerate ground state solutions $\phi_0(x)$ where one is localized around one potential minimum, and the other around the second minimum, i.e., the self-trapping phenomenon is observed. This has been rigorously proven for the double-well potential in the form of Dirac-delta wells, $V\propto -\delta(x)-\delta(x+\pi)$ \cite{Jackson2004}, and also demonstrated experimentally and theoretically in many different double-well potentials, e.g., see analytical solutions in \cite{Mahmud2002}. The proof of the self-trapping phenomenon implies proof of the absolute stability of the corresponding discrete time crystals, as the mapping is mathematically exact. In the following, instead of illustrating the formation of the discrete time crystals in the rotating Dirac-delta wells, we consider an example that can readily be realized in the laboratory \cite{SM}, i.e., when $V=\lambda \cos(2x)$. For sufficiently negative $g_0N$, the ground state reveals the self-trapping phenomenon, where we illustrate the spontaneous symmetry breaking in detail.

In the presence of the external potential ($\lambda\ne 0$) and when the attractive interaction is very small, and in the thermodynamic limit, the ground state of the system (\ref{ll2}) is a BEC where all bosons occupy the single-particle ground state $\phi_0(x)$ which is a balanced superposition of two wave-packets localized in each well of the external potential in (\ref{ll2}) \cite{Pethick2002}. The width of the wave-packets can be estimated by employing the harmonic oscillator approximation for the potential wells and it reads $\sigma\approx1/\sqrt{\Omega}$, where the frequency of the harmonic oscillator $\Omega=2\sqrt{\lambda}$. In the presence of the attractive interactions ($g_0<0$) and increasing their strength, we enter a self-trapping regime for bosons where the mean-field lowest energy solutions are degenerate and each of them can be approximated by all bosons occupying a single wave-packet localized in one well of the external potential \cite{Pethick2002,Jackson2004}. The space translation symmetry of the Hamiltonian (\ref{ll2}) is spontaneously broken and the self-trapped states live forever in the thermodynamic limit. Increasing further the strength of the attractive interactions, we enter the bright soliton regime \cite{Pethick2002}, i.e., the wave-packets localized in the potential wells start shrinking due to strong attractive interactions and resembling the bright soliton solutions. Thus, we can observe three regimes: (i) weakly-interacting regime with no spontaneous symmetry breaking, (ii) moderate-interaction regime and spontaneous breaking of the space translation symmetry of (\ref{ll2}), and (iii) strong-interaction regime where the spontaneous breaking of the symmetry corresponds to the formation of  bright soliton wavepackets of width $\xi=2/
|g_0(N-1)|<\sigma$, the width of the single-particle ground state in the potential well \cite{Pethick2002}. 

\begin{figure}[t]
\centering
\includegraphics[width=0.95\columnwidth]{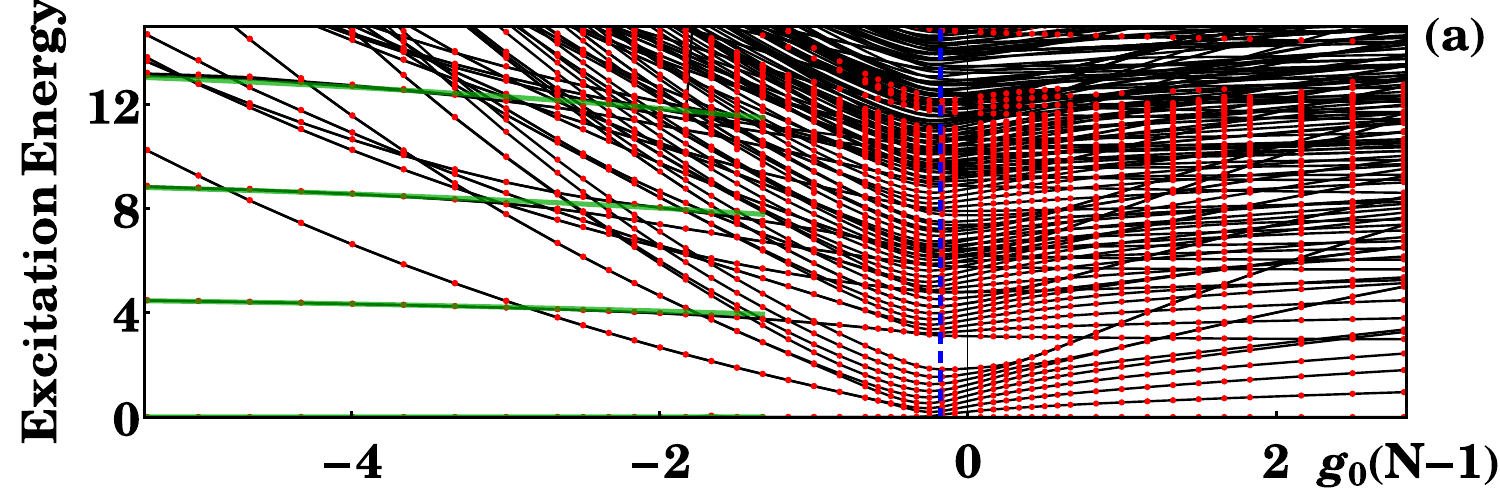}
\includegraphics[width=0.42\columnwidth]{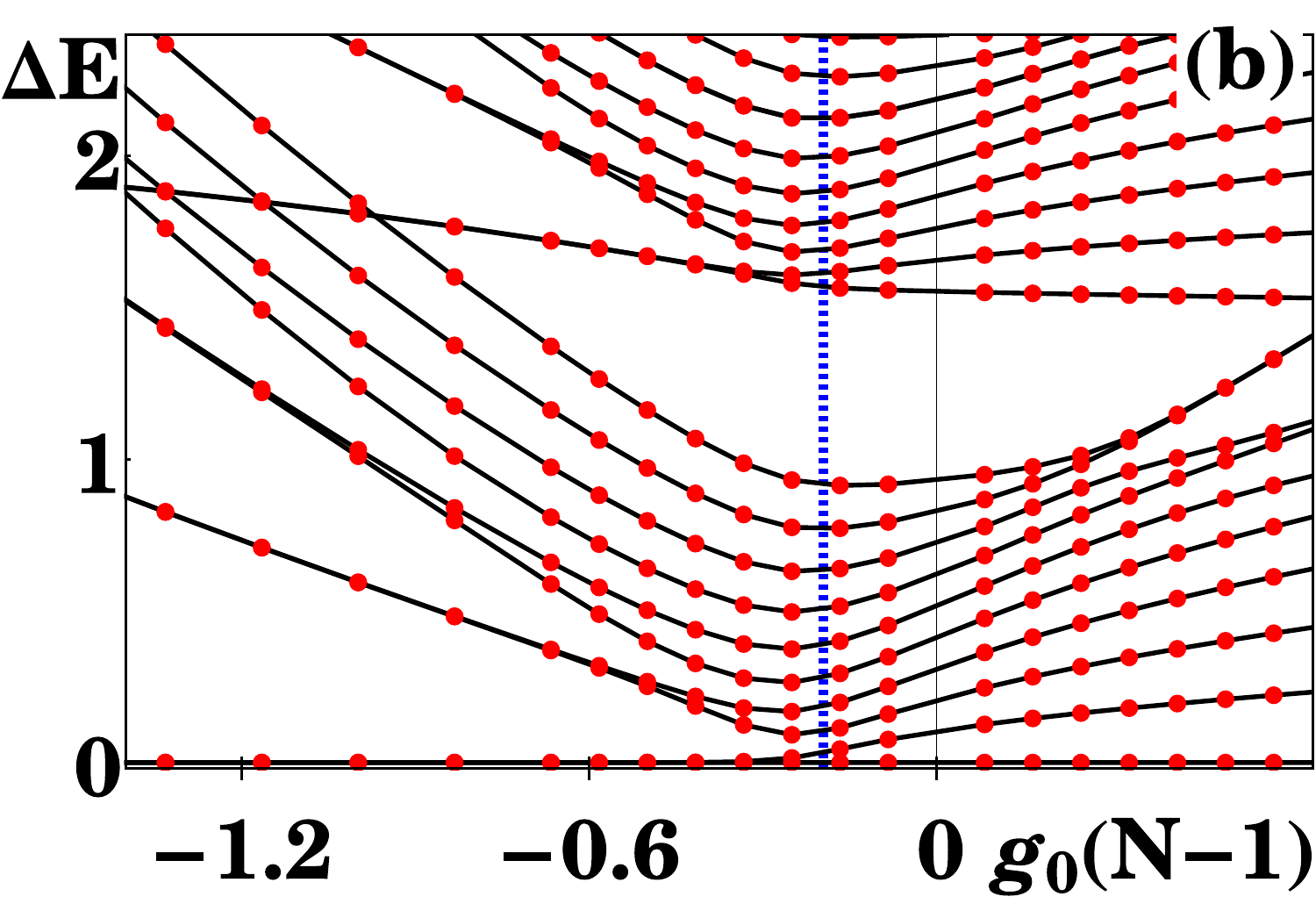}
\includegraphics[width=0.47\columnwidth]{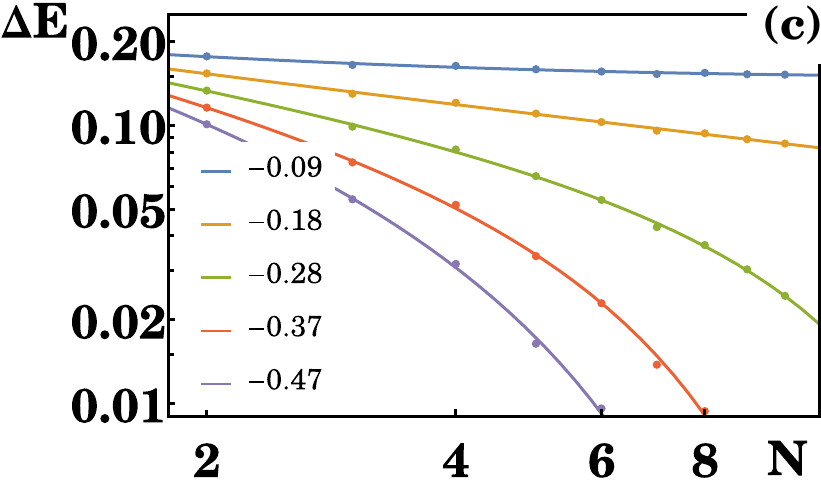}
\caption{
(a) Solid black lines show the excitation spectrum of (\ref{ll2}), i.e., eigenenergies minus the ground state energy, for $N=9$ and $\lambda=1.5$. Vertical dashed line indicates the critical value of $g_0(N-1)$ for the quantum phase transition to the discrete time crystal regime obtained for $N\rightarrow\infty$ from the results presented in (c). Green lines are related to the ground state level and excited levels of the center of mass of the system \cite{SM} which are doubly degenerate because the bosons can be located in one of the wells of the double-well potential. Red circles are exact quasi-energies of the kicked LL model (cf. (\ref{pert}) with $T=\pi/31$ \cite{footnote2}) which are relevant to time crystal states and which are perfectly reproduced by the spectrum of (\ref{ll2}). (b) is an enlargement of (a) in the vicinity of the critical point. (c) Log-log plots of the difference, $\Delta E$ between the lowest eigenenergies of (\ref{ll2}) vs. $N$ for different fixed values of $g_0(N-1)$ as indicated in the figure. The critical value of $g_0(N-1)\approx -0.18$ corresponds to an algebraic decrease of $\Delta E$ with $N$ and this agrees with the two-mode prediction for $N\rightarrow\infty$ \cite{SM}. For weaker interactions $\Delta E$ approaches a constant value, while for stronger interactions, $\Delta E$ decreases exponentially with $N$.
}
\label{spectrum_ll2}
\end{figure}

In order to diagonalize the Hamiltonian (\ref{ll2}), we employ the eigenbasis of the undriven Lieb-Liniger model which can be obtained with the help of the Bethe ansatz \cite{KorepinBook1993,gaudin_2014,SM}. Diagonalization of the Hamiltonian matrix yields the energy spectrum of (\ref{ll2}) which is depicted in Fig.~\ref{spectrum_ll2} as a function of $g_0(N-1)$ for $N=9$ and $\lambda=1.5$. 
Let us first focus on the strongest interactions presented in Fig.~\ref{spectrum_ll2}(a) which are in the regime (iii), where the width $\xi$ of the bright soliton becomes smaller than the width $\sigma$ of the single-particle ground state in the potential well, i.e., when $|g_0(N-1)|\gtrsim 2\sqrt{2}\lambda^{1/4}=3.13$ for $\lambda=1.5$. Without the external potential ($\lambda=0$), the lowest energy solution within the mean-field approach would represent a bright soliton, $\phi_0(x-q)=\cosh^{-1}[(x-q)/\xi]/\sqrt{2\xi}$, which is occupied by all bosons and which can be localized at any point $q$ on the ring \cite{Castin_LesHouches}. In the presence of the double-well potential ($\lambda\ne0$) and for $\xi < \sigma$, there are two possible ground state locations of the soliton at the bottoms of the potential wells, i.e., $q=\pi/2$ or $3\pi/2$. Low-energy excitations of the bright soliton correspond to excitations of its center of mass which are depicted by the green lines in Fig.~\ref{spectrum_ll2}(a) \cite{SM}.
When we decrease the strength of the attractive interactions, we lose such a {\it single-body} character of the low energy spectrum which takes place for $\xi\gtrsim \sigma$. Then, low-energy excitations lead to quantum depletion of a BEC localized in one of the potential wells and transfer of bosons to the other well \cite{Zin2008,Ribeiro2008,Oles2010}. This moderate-interaction regime corresponds to the self-trapping of a BEC where bosons prefer to localize in one of the potential wells but do not form a bound state like in the bright soliton case. The self-trapping properties are observed in the ground and excited eigenstates up to the so-called symmetry breaking edge, i.e., the corresponding excitation energy is proportional to $N$ and the number of states which break the symmetry is extensive which is crucial in order to call discrete time crystals a new phase \cite{SM}. 

One can ask how weak the attractive interactions should be in order to recover the space translation symmetry of (\ref{ll2})? The critical interaction strength for the phase transition between the symmetry-broken and symmetry-preserving phases can be estimated by means of the two-mode approach because the interactions in this regime are weak and not able to modify the shape of the single-particle wave-packets localized in the potential wells which are used in the two-mode approximation \cite{Zin2008,Sacha2015PRA,SM}. Indeed, the two-mode prediction agrees with the numerical results shown in Fig.~\ref{spectrum_ll2}(c) where at the critical value of $g_0(N-1)\approx -0.18$, the energy gap between the lowest energy eigenstates decreases algebraically with $N$. For stronger interactions the gap decreases exponentially while for weaker interactions it approaches a constant value \cite{Sacha2015PRA}. Thus, if the interactions are sufficiently strong and $N\rightarrow\infty$, the symmetry-preserving eigenstates are degenerate and their superpositions form symmetry-broken eigenstates which live forever.

\begin{figure}[t]
\includegraphics[width=0.99\columnwidth]{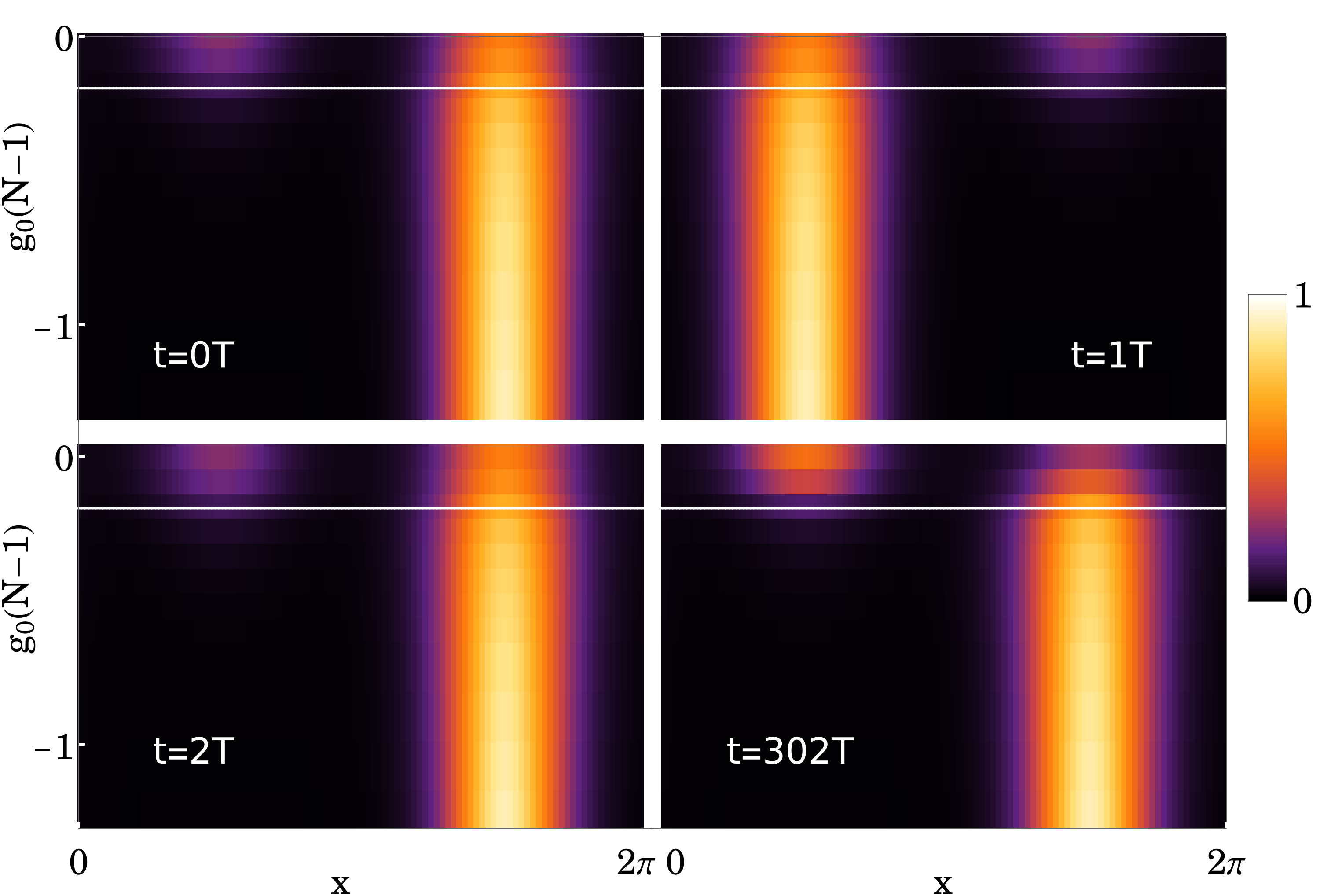}
\caption{
Single-particle probability densities corresponding to superpositions of the two lowest energy eigenstates of (\ref{ll2}) plotted in the lab frame for different moments of time as indicated in the panels for $N=9$. In each panel, collections of the densities for different interaction strength $g_0(N-1)$ are presented. Solid white lines indicate the critical interaction strength $g_0(N-1)\approx-0.18$ for the transition to the discrete time crystal regime when $N\rightarrow\infty$. Densities above these lines reveal a decay of the $2T$-periodic evolution due to tunneling --- for non-interacting bosons, complete tunneling takes place at $t\approx 302T$. Densities below the solid white lines show discrete time crystal evolution for $N\rightarrow\infty$. The other parameters are the same as in Fig.~\ref{spectrum_ll2}, i.e., $\lambda=1.5$ and $T=\pi/31$ \cite{footnote2}.
}
\label{density}
\end{figure}

When we return to the laboratory frame, the symmetry-broken eigenstates of (\ref{ll2}) will evolve with a period of $2T$, demonstrating discrete time crystals with absolute stability. In Fig.~\ref{density} we present the time evolution of superpositions of the lowest symmetry-preserving eigenstates of (\ref{ll2}) for different interaction strengths. At $g_0(N-1)\approx -0.18$ and for $N\rightarrow\infty$, the quantum phase transition to the discrete time crystal regime occurs where the superpositions evolve with the period $2T$ --- for $N=9$ the ground state level becomes practically degenerate for stronger interactions, i.e., $g_0(N-1)\approx -0.3$. At slightly stronger interactions, excited levels also become practically doubly degenerate, indicating that with increasing $N$ there is an extensive number of states that reveal time crystal behavior.  At $g_0(N-1)$ around $-3$, there is a crossover to the bright soliton regime where low-energy excitations have {\it single-body} character because bosons form a bound bright soliton state \cite{Dziarmaga2004,Weiss2009,Sacha2009}.

It is important to emphasize that the spontaneous emergence of the time crystal and its evolution with the period $2T$ is not a trivial observation of a time-independent problem in a rotating frame. If the interactions were too weak to induce spontaneous symmetry breaking, when returning from the rotating frame to the laboratory frame, we would observe periodic evolution with the period $T$ because such a period corresponds to the time translational symmetry  of the Hamiltonian (\ref{ll1}). Only when the interactions lead to symmetry breaking, we observe the emergence of the time crystal and its evolution with the period $2T$.

Having analyzed the system (\ref{ll1}), we will now switch to a system with a periodically kicked potential, which can also be experimentally implemented and has recently garnered significant interest \cite{Cao2022,SeeToh2022}. Although the stability of discrete time crystals in the kicked LL model remains an open question, this model exhibits a range of different time and space symmetry breaking. Additionally, it provides intriguing opportunities to investigate the impact of a reservoir of a rotating thermal cloud on the stability of phases with broken symmetries. Let us exchange the time-periodic perturbation in the Hamiltonian (\ref{ll1}) with
\be
H_1=\lambda T\sum_{i=1}^N \cos(2x_i)\sum_{m=-\infty}^{+\infty}\delta(t-mT),
\label{pert}
\ee 
which describes periodic kicking of the particles with the period $T=2\pi/\omega$. When we perform the same unitary transformation to the moving frame as previously, i.e., $U_t$, and a similar shift of the momenta $U_p$, then using the rotating-wave approximation or the Magnus expansion we can obtain an effective Hamiltonian identical to (\ref{ll2}), see \cite{SM,Blanes2010}. 

We have already analyzed the system (\ref{ll2}); thus, in the present case of the time-periodic kicking (\ref{pert}), we have to demonstrate only that the low-energy eigenstates of (\ref{ll2}) reproduce well the relevant exact eigenstates of the Floquet Hamiltonian which are also eigenstates of the Floquet evolution operator. The Floquet evolution operator is the evolution operator of the system over a single driving period which in the case of the time-periodic kicking (\ref{pert}) reads
\be
U(T)=e^{-iH_{\rm LL}T}\;e^{-i\lambda T\sum_{i=1}^N \cos(2x_i)}.
\label{uf}
\ee
This unitary operator can be diagonalized in the eigenbasis of the LL Hamiltonian and knowing its eigenphases $\phi_n$, we can calculate the quasi-energies of the system, $E_n=-\phi_n/T$. The obtained exact quasi-energies, which are relevant to discrete time crystal states, are also shown in Fig.~\ref{spectrum_ll2} and they are perfectly reproduced by the low-lying spectrum of (\ref{ll2}). We have chosen the same $\lambda$ as previously and consequently we expect exactly the same spectrum as in the case of the Hamiltonian (\ref{ll1}), and indeed the obtained quasi-energies are indistinguishable in the plot.

Both the system (\ref{ll1}) and the kicked LL model (\ref{pert}) can be described using the Floquet formalism, where all physically relevant quasi-energies lie in a single Floquet zone. For small values of $N$, Floquet states can be obtained numerically. However, very high-order terms, which are neglected in numerical calculations, may introduce tiny couplings between Floquet states with similar quasi-energies that can lead to the decay of discrete time crystals as $t\rightarrow\infty$, especially when we first take the $N\rightarrow\infty$ limit. In the case of the system (\ref{ll1}), we know that quasi-energies corresponding to the discrete time crystal can be unfolded and they correspond to the low-energy spectrum of the time-independent Hamiltonian (\ref{ll2}). Furthermore, we have a guarantee that they are not coupled to any other Floquet states in any order. In the case of the kicked LL model, such absolute stability of the discrete time crystals can not be guaranteed. The discrete time crystal in (\ref{ll1}) has been designed so that the perturbation contains only one harmonic responsible for the formation of the crystal. In the case of the kicked LL model, there are many other harmonics whose influence can be reduced but cannot be fully eliminated.

So far we have analyzed the Floquet states that break both discrete space and time translational symmetry in the periodically kicked LL model by switching to the frame moving with the frequency $\omega/2$ by means of the unitary transformation $U_t$. Let us show that the periodically kicked LL model (\ref{pert}) can also support states that spontaneously break the discrete space translation symmetry without breaking the discrete time translation symmetry, which does not exist in the corresponding LL model driven by the rotating potential. To investigate these states, we can also switch to the moving frame but with the frequency $\omega$, and we still obtain the same effective Hamiltonian (\ref{ll2}) \cite{SM}. Then, however, eigenstates of (\ref{ll2}) that reveal spontaneous breaking of the discrete space translation symmetry, do not break the discrete time translation symmetry. Indeed, when we return to the laboratory frame, all eigenstates of (\ref{ll2}) evolve with the driving period $T$. 

The periodically kicked LL model is also attractive in studying the time crystal behavior in contact with a thermal bath. In the case of a rotating potential (\ref{ll1}), the discrete time crystals should be accessible as a thermal equilibrium state in contact with a reservoir of rotating thermal atoms. In the case of (\ref{pert}), the situation becomes more intriguing, as in the rotating frame, the time-independent Hamiltonian (\ref{ll2}) is only an approximation. Cooling of atoms in the presence of a rotating thermal cloud has been realized in experiments demonstrating vortices in a BEC \cite{Pethick2002}. So, new opportunities for theoretical and experimental investigations of the stability of discrete time crystals in contact with a thermal reservoir are opening up.

To summarize, we have analyzed interacting bosons on a ring with various periodic perturbations which turns out to be a suitable system for realization of discrete time crystals. Most importantly, this system can reveal discrete time crystals whose stability can be proven by an exact mapping of the discrete time crystal states to low-lying eigenstates of a time-independent Hamiltonian. The periodically driven bosons on a ring can also reveal big discrete time crystals and condensed matter in time \cite{Giergiel2018a,sacha15a,Hannaford2022,Guo2013,GuoBook2021,Giergiel2022}. For example, if we choose $\cos(sx_i - \omega t)$ as the potential in (\ref{ll1}), then spontaneously emerging discrete time crystals can evolve with a period $s$ times longer than the driving period, where $s \gg 1$.

This research was funded by the National Science Centre, Poland, Projects No. 2018/31/B/ST2/00349 (K.G.) and No. 2021/42/A/ST2/00017 (K.S.) and the Australian Research Council (ARC), Project DP190100815. K.G. acknowledges the support of the Polish National Agency for Academic Exchange Bekker Programme (BPN/BEK/2021/1/00339). J.W. acknowledges the support of ARC Discovery Program FT230100229. Numerical work was performed on the OzSTAR national facility at Swinburne University of Technology. The OzSTAR program receives funding in part from the Astronomy National Collaborative Research Infrastructure Strategy (NCRIS) allocation provided by the Australian Government, and from the Victorian Higher Education State Investment Fund (VHESIF) provided by the Victorian Government.

%\bibliography{ll_Ref_v19}

%%%%%%%%%%%%%%%%%%%%%%%%%%%%%%%%%%%%%%%%%%%%%%%%%%%%%%%%%%%%%%%%%%%%%%%%%%%%%%%%%%%%
% SUPPLEMENTAL MATERIAL
%%%%%%%%%%%%%%%%%%%%%%%%%%%%%%%%%%%%%%%%%%%%%%%%%%%%%%%%%%%%%%%%%%%%%%%%%%%%%%%%%%%%

%\begin{widetext}
\begin{center}
{\bf \Large Supplemental Material\\}
%{\bf \Large SUPPLEMENTAL MATERIAL\\}
\end{center}
%\end{widetext}

%%%%%%%%%% Merge with supplemental materials %%%%%%%%%%
%%%%%%%%%% Prefix a "S" to all equations, figures, tables and reset the counter %%%%%%%%%%
\setcounter{equation}{0}
\setcounter{figure}{0}
\setcounter{table}{0}
\makeatletter
\renewcommand{\theequation}{S\arabic{equation}}
\renewcommand{\thefigure}{S\arabic{figure}}
%\renewcommand{\bibnumfmt}[1]{[S#1]}
%\renewcommand{\citenumfont}[1]{S#1}
%%%%%%%%%% Prefix a "S" to all equations, figures, tables and reset the counter %%%%%%%%%%

In this Supplemental Material, we provide details regarding the description and analysis of the discrete time crystals presented in the Letter. We focus on the example of a rotating potential, which we use to illustrate the formation of the discrete time crystals discussed in the Letter. We begin with the more general case where the rotating potential takes the form $\cos(sx-\omega t)$, where $s$ is an arbitrary integer. When $s \gg 1$, big discrete time crystals can emerge, evolving with a period that is $s$ times longer than the period of the periodic perturbation~\cite{Hannaford2022}.

In Sec.~\ref{rotatingLL} we present: details of the transformations used in the Letter to describe the Lieb-Liniger(LL) model driven by a rotating potential, analysis of consequences of breaking of space translation symmetry in the moving frame on time translation symmetry in the laboratory frame, two-mode analysis of the formation of discrete time crystals, a short description of the Bethe ansatz approach, analysis of the bright soliton regime in the LL model and an example of experimentally attainable parameters 

In Sec.~\ref{KICKEDLL}  we present: a derivation of the effective Hamiltonian of the resonantly kicked LL model and a demonstration that in the kicked LL model space translation symmetry can be spontaneously broken without breaking of time translation symmetry.

\section{Lieb-Liniger model driven by a rotating lattice potential}
\label{rotatingLL}

Let us consider $N$ bosons with contact interactions on a ring with circumference $2\pi$ which are periodically driven by a rotating potential $\cos(sx-\omega t)$ where $s$ is an integer number. The system reduces to the Lieb-Linger (LL) Hamiltonian $H_{\rm LL}$ with an additional time-periodic drive
\bea
H&=&H_{\rm LL}+\lambda\sum_{i=1}^N \cos(sx_i-\omega t),
\label{Sll1} \\
H_{\rm LL}&=&\sum_{i=1}^N\frac{p_i^2}{2}+g_0\sum_{i<j}^N\delta(x_i-x_j),
\label{Sll1a}
\eea
where we use $R$ and $\hbar^2/mR^2$ for the length and energy units, respectively, where $R$ is the ring radius and $m$ the mass of the bosons. The interaction strength $g_0=2mR\omega_\perp a_s/\hbar$, where $\omega_\perp$ is the frequency of the harmonic transverse confinement and $a_s$ is the atomic s-wave scattering length.
The system possesses discrete time and space translation symmetries, i.e., the Hamiltonian (\ref{Sll1}) does not change if $t\rightarrow t+T$ (where $T=2\pi/\omega$) or all $x_i\rightarrow x_i+2\pi/s$.

We have chosen the potential of the form of $\cos(sx-\omega t)$ to drive the LL model but one can choose many different potentials $V(sx-\omega t)$ and similar phenomena that we consider in the Letter can be realized.

\subsubsection*{Note on conventions}

In the Bethe ansatz literature the following convention is often used \cite{KorepinBook1993,gaudin_2014}
\bea
H'&=&H'_{\rm LL}+\lambda'\sum_{i=1}^N \cos(sx_i-\omega t),
\\
H'_{\rm LL}&=&\sum_{i=1}^N p_i^2+2c\sum_{i<j}^N\delta(x_i-x_j).
\eea
To switch to the convention used in (\ref{Sll1})-(\ref{Sll1a}), the following scaling has to be applied
\bea
c=g_0,\\
\lambda'=2\lambda,\\
T'=T/2,\\
H'=2H.
\eea

\subsection{Transformation to moving frame: Hamiltonian and quantum state}
\label{SecIA}

The model Hamiltonian (\ref{Sll1}) becomes easier to investigate in the moving frame of
the rotating potential, where we first perform a time-dependent unitary
transformation $U_{t}=\exp (i\sum_{j}p_{j}\omega t/s)$, leading to a shift
in the positions, $x_{i}\rightarrow x_{i}+\omega t/s$, and next a second
time-independent unitary transformation $U_{p}=\exp \left(
-i\sum_{j}x_{j}\omega /s\right) $, leading to a shift in the momenta, $%
p_{i}\rightarrow p_{i}+\omega /s$. Under these transformations, the
time-dependent quantum Liouville-von Neumann equation in the lab frame $%
i\partial _{t}\rho =[H,\rho ]$ becomes $i\partial _{t}\tilde{\rho}=[\tilde{H}%
,\tilde{\rho}]$ in the moving frame, where $\tilde{\rho}=U_{p}U_{t}\rho
U_{t}^{\dagger }U_{p}^{\dagger }$ and $\rho $ are notations for density
operators in the moving frame and lab frame, respectively. As will be shown
below, we end up with the following exact time-independent Hamiltonian 
\begin{equation}
\tilde{H}=\sum_{i=1}^{N}\left[ \frac{p_{i}^{2}}{2}+\lambda
\cos(sx_i)\right] +g_{0}\sum_{i<j}^{N}\delta (x_{i}-x_{j}),
\label{Sll2}
\end{equation}%
where a constant term has been omitted. This new Hamiltonian possesses the
same discrete space translation symmetry as the Hamiltonian (\ref{Sll1}).

The quantities $\tilde{\rho}(\vec{x},\vec{x}^{\prime };t)$ and $\rho (\vec{x}%
,\vec{x}^{\prime };t)$ are matrix elements of $\tilde{\rho}$ and $\rho $ for
position eigenstates. Thus, $\rho (\vec{x},\vec{x}^{\prime };t)=\left\langle 
\vec{x}\right\vert \,\rho _{t}\,\left\vert \vec{x}^{\prime }\right\rangle $,
with $\vec{x}\equiv \{x_{1},x_{2},...,x_{N}\}$ and the subscript $t$
specifies the density operator at time $t$. The diagonal elements of $\rho (%
\vec{x},\vec{x}^{\prime };t)$ are related to the position single-particle probability
density in the laboratory frame $P(x,t)$ via 
\begin{equation}
P(x,t)=\int dx_{2}dx_{3}...dx_{n}\,\rho (\vec{x},\vec{x},;t),
\label{Eq.PosProbDensity}
\end{equation}%
where $\vec{x}=\{x,x_{2},x_{3},...x_{N}\}$. \medskip

To show that the moving frame density operator $\tilde{\rho}$ satisfies the
equation $i\partial _{t}\tilde{\rho}=[\tilde{H},\tilde{\rho}]$ involving the
transformed Hamiltonian $\tilde{H}$ we begin with the time-dependent
Schr\"odinger equation $(H-i\partial _{t})\,\left\vert \Psi \right\rangle =0$
for a quantum state $\left\vert \Psi \right\rangle $ in the laboratory
frame, and $H$ given by (\ref{Sll1}). The quantum state satisfies
periodicity conditions for the ring of length $2\pi $; thus $\left\langle 
\vec{x}+2\pi |\Psi \right\rangle =\left\langle \vec{x}|\Psi \right\rangle $, where $\vec x+2\pi\equiv\{x_1+2\pi,x_2+2\pi,\ldots,x_N+2\pi\}$.

We first apply the unitary transformation $U_{t}$. This is straight-forward
using the well-known identity $\exp (S)\;\Omega \;\exp (-S)=\Omega
+[S,\Omega ]+(1/2)[S,[S,\Omega ]]+\ldots$ and the commutation rules $%
[x_{j'},p_{j}]=i\delta _{j'j}$ to show that $U_{t}\,x_{i}\,U_{t}^{-1}=x_{i}+%
\omega t/s$ and $U_{t}\,p_{i}\,U_{t}^{-1}=p_{i}$, so that 
\begin{equation}
U_{t}\,H\,U_{t}^{-1}=\sum_{i=1}^{N}\frac{p_{i}^{2}}{2}+g_{0}\sum_{i<j}^{N}%
\delta (x_{i}-x_{j})+\lambda \sum_{i=1}^{N}\cos(sx_{i}).
\end{equation}%
We then have 
\begin{equation}
U_{t}\,H\,U_{t}^{-1}\left( U_{t}\,\left\vert \Psi \right\rangle \right)
-iU_{t}\,\frac{\partial }{\partial t}\left( U_{t}^{-1}\;U_{t}\,\left\vert
\Psi \right\rangle \right) =0.  \label{Eq.UtResult}
\end{equation}

\begin{widetext}
Now 
\begin{eqnarray}
\frac{\partial }{\partial t}\left( U_{t}^{-1}U_{t}\,\left\vert \Psi
\right\rangle \right) &=&\frac{\partial }{\partial t}\left(
U_{t}^{-1}\right) \times \left( U_{t}\,\left\vert \Psi \right\rangle \right)
+\left( U_{t}^{-1}\right) \times \frac{\partial }{\partial t}\left(
U_{t}\,\left\vert \Psi \right\rangle \right)  \notag \\
&=&(\frac{-i\omega }{s}\sum_{j}p_{j})\left( U_{t}^{-1}\right) \times \left(
U_{t}\,\left\vert \Psi \right\rangle \right) +\left( U_{t}^{-1}\right)
\times \frac{\partial }{\partial t}\left( U_{t}\,\left\vert \Psi
\right\rangle \right),
\end{eqnarray}%
since it is straight-forward to show that $\frac{\partial }{\partial t}\left(
U_{t}^{-1}\right) =(\frac{-i\omega }{s}\sum_{j}p_{j})\left(
U_{t}^{-1}\right) $. As $U_{t}$ commutes with $p_{j}$, substituting
into (\ref{Eq.UtResult}) then gives 
\begin{eqnarray}
\left( \sum_{i=1}^{N}\frac{p_{i}^{2}}{2}+g_{0}\sum_{i<j}^{N}\delta
(x_{i}-x_{j})+\lambda \sum_{i=1}^{N}\cos(sx_{i})-\frac{\omega }{s}%
\sum_{i}p_{i}-i\frac{\partial }{\partial t}\right) \left( U_{t}\,\left\vert
\Psi \right\rangle \right) &=&0,
\end{eqnarray}%
and consequently
\begin{eqnarray}
 \left( \sum_{i=1}^{N}\frac{(p_{i}-\frac{\omega }{s})^{2}}{2}%
+g_{0}\sum_{i<j}^{N}\delta (x_{i}-x_{j})+\lambda \sum_{i=1}^{N}\cos(sx_{i})-%
\frac{N\omega ^{2}}{2s^{2}}-i\frac{\partial }{\partial t}\right) \left(
U_{t}\,\left\vert \Psi \right\rangle \right) &=&0.  \label{Eq.TDSEforUtPsi}
\end{eqnarray}%
Note the momentum off-set $\omega/s$ when only $U_{t}$ is applied.

Secondly, we apply the unitary transformation $U_{p}$. It is straight-forward
to show that $U_{p}\,x_{i}\,U_{p}^{-1}=x_{i}$ and $U_{p}\,p_{i}%
\,U_{p}^{-1}=p_{i}+\omega /s$, so that from the last equation we obtain
\begin{eqnarray}
U_{p}\left( \sum_{i=1}^{N}\frac{(p_{i}-\frac{\omega }{s})^{2}}{2}%
+g_{0}\sum_{i<j}^{N}\delta (x_{i}-x_{j})+\lambda \sum_{i=1}^{N}\cos(sx_{i})-%
\frac{N\omega ^{2}}{2s^{2}}\right) U_{p}^{-1}\left( U_{p}U_{t}\,\left\vert
\Psi \right\rangle \right) -iU_{p}\frac{\partial }{\partial t}\left(
U_{p}^{-1}U_{p}U_{t}\,\left\vert \Psi \right\rangle \right) &=&0,
\cr &&
\end{eqnarray}%
and finally
\begin{eqnarray}
\left( \sum_{i=1}^{N}\frac{p_{i}{}^{2}}{2}+g_{0}\sum_{i<j}^{N}\delta
(x_{i}-x_{j})+\lambda \sum_{i=1}^{N}\cos(sx_{i})-\frac{N\omega ^{2}}{2s^{2}}-i%
\frac{\partial }{\partial t}\right) \left( U_{p}U_{t}\,\left\vert \Psi
\right\rangle \right) &=&0,  \label{Eq.TDSEforUpUtPsi}
\end{eqnarray}%
noting that $U_{p}$ is not time dependent, so $\frac{\partial }{\partial t}%
\left( U_{p}^{-1}\right) =0$. Apart from the constant $-\frac{N\omega ^{2}}{%
2s^{2}}$, the Hamiltonian in (\ref{Eq.TDSEforUpUtPsi}) is the same as (\ref{Sll2}). The quantum state in the moving
frame is $| \tilde{\Psi }\rangle =U_{p}U_{t}\,\left\vert
\Psi \right\rangle $.

The generalization to mixed states, which in the laboratory frame would be
of the form $\rho =\sum_{\Psi }P_{\Psi }\vert \Psi\rangle \langle \Psi \vert $, is obvious, and results in the
Liouville-von Neumann equation (described above) involving the
time-independent Hamiltonian (\ref{Sll2}).

\subsection{Position probability density in moving and laboratory frames}
\label{SecIB}

Since the Hamiltonian (\ref{Sll2}) is time-independent, there exist
well-defined energy eigenstates and equilibrium states, so the density
matrix elements in the moving frame $\tilde{\rho}(\vec{x},\vec{x}^{\prime })$ can be time independent. The corresponding matrix elements in the lab
frame at time $t$ are given by 
\begin{eqnarray}
\rho (\vec{x},\vec{x}^{\prime };t) &=&\left\langle \vec{x}\right\vert
\,U_{t}^{\dagger }U_{p}^{\dagger }\;\tilde{\rho}\;U_{p}U_{t}\,\left\vert 
\vec{x}^{\prime }\right\rangle  \notag =\left\langle \vec{x}-\omega t/s\right\vert U_{p}^{\dagger }\;\tilde{\rho}\;
U_{p}\left\vert \vec{x}^{\prime }-\omega t/s\right\rangle  \notag 
=\left\langle \vec{x}-\omega
t/s\right\vert\tilde{\rho}\left\vert \vec{x}^{\prime }-\omega
t/s\right\rangle  \notag e^{i\varphi (\vec{x},\vec{x}^{\prime })}\\
&=&\tilde{\rho}\left(\vec{x}-\frac{\omega t}{s},\vec{x}^{\prime }-\frac{\omega
t}{s}\right)e^{i\varphi (\vec{x},\vec{x}^{\prime })},  \label{Eq.DensityMatrixRelns}
\end{eqnarray}%
where $\varphi (\vec{x},\vec{x}^{\prime })=\sum_{i}
(x_{i}-x_{i}^{\prime })\omega/s$ is a phase that only depends on relative
distances and originates from the operator $U_{p}$. It does not depend on the
quantum state. Using (\ref{Eq.PosProbDensity}) we then see that the single-particle
probability densities $P(x,t)$ and $\tilde{P}(x,t)$ in  the
laboratory and moving frames are related via 
\begin{equation}
P(x,t)=\tilde{P}(x-\omega t/s).
\label{Eq.PositProbDensRelations}
\end{equation}

Suppose in the moving frame a time-independent $\tilde{\rho }(\vec{x},\vec{x}^{\prime })$ has a spatial
periodicity $2\pi$, i.e., $\tilde{\rho}\left(\vec{x}-2\pi,\vec{x}
^{\prime }-2\pi\right)=\tilde{\rho}\left(\vec{x},\vec{x}
^{\prime }\right)$.
Let us consider what happens in the laboratory frame at time $t+sT$
\begin{eqnarray}
\rho (\vec{x},\;\vec{x}^{\prime };\;t+sT)&=&\tilde{\rho}\left(\vec{x}
-2\pi -\frac{\omega t}{s},\;\vec{x}^{\prime }-2\pi -\frac{\omega t}{s}\right)e^{i\varphi (\vec{x},\vec{x}^{\prime })}
\cr 
&=&\tilde{\rho}\left(\vec{x}
-\frac{\omega t}{s},\;\vec{x}^{\prime } -\frac{\omega t}{s}\right)e^{i\varphi (\vec{x},\vec{x}^{\prime })}
\cr 
&=& \rho (\vec{x},\;\vec{x}^{\prime };\;t).
\end{eqnarray}
So, in the laboratory frame, the density
matrix elements $\rho (\vec{x},\vec{x}^{\prime };t)$ have time periodicity $%
sT$. From (\ref{Eq.PositProbDensRelations}) the same applies to the single-particle
probability density in the laboratory frame.

There could of course be other possible space periodicities for the time
independent moving frame density matrix elements, such as $2\pi/s$, since
there are $s$ identical potential wells between $0$ and $2\pi $ and the Hamiltonian (\ref{Sll2}) is invariant under translation of all bosons in space by $2\pi/s$. But suppose
this is not the case, i.e.,
\begin{equation}
\tilde{\rho}\left(\vec{x}-\frac{2\pi}{s},\;\vec{x}^{\prime }-\frac{2\pi}{s}\right)\neq \tilde{\rho}\left(\vec{x},\;\vec{x}^{\prime }\right).
\end{equation}
Now consider what happens at time $t+T$ in the laboratory frame. Similarly to before we would have 
\begin{eqnarray}
\rho (\vec{x},\vec{x}^{\prime };\;t+T)&=&
\tilde{\rho}\left(\vec{x}-\frac{\omega (t+T)}{s},\;\vec{x}^{\prime }-\frac{\omega
(t+T)}{s}\right)e^{i\varphi (\vec{x},\vec{x}^{\prime })}
\cr
&=&\tilde{\rho}\left(\vec{x}
-\frac{2\pi}{s}-\frac{\omega t}{s},\;\vec{x}^{\prime }-\frac{2\pi}{s}-\frac{\omega t}{s}\right)e^{i\varphi (\vec{x},\vec{x}^{\prime })}
\cr 
&\ne& \tilde{\rho}\left(\vec{x}
-\frac{\omega t}{s},\;\vec{x}^{\prime }-\frac{\omega t}{s}\right)e^{i\varphi (\vec{x},\vec{x}^{\prime })},
\end{eqnarray}
\end{widetext}
due to the previous inequality and consequently $\rho (\vec{x},\vec{x}^{\prime};t+T)\neq \rho (\vec{x},\vec{x}^{\prime };t)$. 

Thus, the failure of space periodicity $2\pi /s$ in the moving
frame results in the failure to have time periodicity $T$ in the laboratory
frame. Conversely, if in the moving frame $\tilde{\rho }(\vec{x},\vec{x}
^{\prime })$ has a spatial periodicity $2\pi /s$, then $%
\rho (\vec{x},\vec{x}^{\prime };t+T)=\rho (\vec{x},\vec{x}^{\prime };t)$ and
the density matrix elements $\rho (\vec{x},\vec{x}^{\prime };t)$ then have
time periodicity $T$, the same as that of the drive in (\ref{Sll1}). We have therefore
shown that spontaneous breaking of the discrete space
translation symmetry of (\ref{Sll2}) corresponds to spontaneous breaking of
the discrete time translation symmetry of the original Hamiltonian (\ref{Sll1}). That is, in the moving frame the symmetry-broken ground state of (\ref{Sll2}) corresponds to a state which evolves with the period $sT$ in the laboratory frame and consequently describes a discrete
time crystal. For a space crystal we have a large number of regular
repetitions in space of an observable measured at any given time, whereas
for the time crystal with $s\gg1$ we have a large number of regular repetitions in time
of an observable measured at any given position. For $s\gg 1$ we deal with a
big discrete time crystal \cite{Giergiel2018a,Hannaford2022}. 

\subsection{Non-integer $\omega/s$ frequency}

Note, that if $\omega/s$ is not an integer number ne can still obtain a result in the form (\ref{Sll2}), but this comes at the cost of the original periodic boundary conditions becoming twisted boundary conditions. 
The exact symmetry of the Lieb-Liniger model are momentum translations fulfilling the periodic condition $\exp(i 2\pi \Delta p)=1$. The transformation $U_p=\exp\left(-i\sum_jx_j\Omega\right)$ can be easily performed up to the largest value for which this condition is fulfilled $\Omega=\omega/s-\left(\omega/s \bmod 1\right)$ resulting in:
\bea
\tilde{H}&=&\sum_{i=1}^{N}\left[ \frac{\left[p_{i}-\left(\omega/s \bmod 1\right)  \right]^{2}}{2}+\lambda\cos(sx_i)\right]\\
&+&g_{0}\sum_{i<j}^{N}\delta (x_{i}-x_{j}).
\label{SllNI}
\eea
This is now a ring that is threaded by a flux smaller than a quantum of momentum. This flux inhibits transport between the minimal points in the potential, due to the destructive interference between clockwise and counter-clockwise paths. This should lead to a symmetry breaking regime appearing at lower interactions value than for the integer case. 

\subsection{Two-mode approximation}
\label{2modeSec}

Let us focus on the $s=2$ case which corresponds to the double-well potential in the Hamiltonian (\ref{Sll2}). If we consider only a single particle, then the lowest energy eigenstates of (\ref{Sll2}) are symmetric and anti-symmetric superpositions of two wavepackets, $w_{1,2}(x)$, localized in the potential wells, i.e., $w_1(x)\pm w_2(x)$. The corresponding eigenenergies are slightly split by 
\be
J=-2\int_0^{2\pi}dx \; w_2^*(x)\left[\frac{p^2}{2}+\lambda\cos(2x)\right]w_1(x),
\ee
which is the tunneling amplitude --- if a particle is initially prepared in $w_1$ (or $w_2$) state, it tunels to the other potential well after time $\pi/J$. If there are $N$ non-interacting bosons ($g_0=0$), then in the Hilbert subspace spanned by the two modes $w_{1,2}$, the Hamiltonian (\ref{Sll2}) reduces to $\tilde H=-J(\hat a_1^\dagger \hat a_2+\hat a_2^\dagger \hat a_1)/2$, where a constant term is omitted and $\hat a_{1,2}$ are the standard bosonic anihilation operators. If the contact interactions are present ($g_0\ne 0$), then in the same Hilbert subspace, the Hamiltonian reads \cite{Pethick2002,SachaTC2020}
\be
\tilde H \approx -\frac{J}{2}(\hat a_1^\dagger \hat a_2+\hat a_2^\dagger \hat a_1)+
\frac{U}{2}\left(\hat a_1^\dagger\hat a_1^\dagger\hat a_1\hat a_1+\hat a_2^\dagger\hat a_2^\dagger\hat a_2\hat a_2\right),
\label{2modeH}
\ee
with $U=U_{11}-2U_{12}$, where 
\be
U_{ij}=g_0\int_0^{2\pi}dx \; |w_i(x)|^2|w_j(x)|^2.
\ee
The two-mode approximation (\ref{2modeH}) is valid provided the interaction energy per particle is much smaller than the energy gap to the next single-particle excited eigenstate which is not included in the Hilbert subspace spanned by the two modes $w_{1,2}$, i.e., provided $NU\ll 2\sqrt{\lambda}$. 

The Hamiltonian (\ref{2modeH}) can be reduced to the form of the Lipkin-Meshkov-Glick model by defining the spin operators, $\hat S_x=(\hat a_1^\dagger \hat a_2+\hat a_2^\dagger \hat a_1)/2$ and $\hat S_z=(\hat a_2^\dagger \hat a_2-\hat a_1^\dagger \hat a_1)/2$. If we use $\hat S_{x,z}=\sum_{i=1}^N\sigma_i^{x,z}/2$, where $\sigma_i^{x,z}$ are the Pauli matrices, the Lipkin-Meshokov-Glick model takes the form of a system of $N$ spin-1/2 particles with all-to-all interactions \cite{Ribeiro2008,SachaTC2020}, 
\be
\tilde H\approx \frac{J}{2}\left(-\sum_{i=1}^N\sigma_i^x+\frac{\gamma}{2N}\sum_{i=1}^N\sum_{j=1}^N\sigma_i^z\sigma_j^z\right),
\label{LMG}
\ee
where 
\be
\gamma=\frac{U(N-1)}{J}.
\ee
In the Lipkin-Meshkov-Glick model, there is a quantum phase transition between the paramagnetic and ferromagnetic phases when the ferromagnetic interactions ($U<0$) between the spins are sufficiently strong. In the two-mode Hamiltonian (\ref{2modeH}) this phase transition corresponds to self-trapping of bosons in one of the modes $w_{1,2}$ (i.e., in one of the potential wells) and consequently spontaneous breaking of the space translation symmetry of the Hamiltonian (\ref{Sll2}) if the attractive interactions ($g_0<0$) between bosons are sufficiently strong \cite{Pethick2002,Jackson2004}.

The critical interaction strength for the spontaneous breaking of the space translation symmetry of the Hamiltonian (\ref{2modeH}) can be obtained by applying the mean-field theory where one looks for the ground state of the system in the form of a product state (i.e., a Bose-Einstein condensate state) 
\be
\tilde\psi(\vec x)=\prod_{i=1}^N\left[a_1w_1(x_i)+a_2w_2(x_i)\right]. 
\label{meanfield}
\ee
The mean-field ground state is determined by the minimal value of the energy functional 
\be
E=NJ\left[-\frac{1}{2}(a_1^* a_2+a_2^* a_1)+
\frac{\gamma}{2}\left(|a_1|^4+|a_2|^4\right)\right],
\ee 
with the constraint $|a_1|^2+|a_2|^2=1$.
For $\gamma>-1$, 
%where
%\be
%\gamma=\frac{U(N-1)}{J},
%\ee
the lowest energy state corresponds to the symmetric superposition of the modes, i.e., $a_1=a_2=1/\sqrt{2}$ in (\ref{meanfield}). However, at $\gamma=-1$, there is a bifurcation and two degenerate symmetry-broken states are born which possess the lowest energy when $\gamma<-1$ 
\be
a_{1,2}=\left(\frac{1\pm \sqrt{1-1/\gamma^2}}{2}\right)^{1/2}.
\ee
Thus, in the mean-field limit (i.e., when $N\rightarrow\infty$ but $\gamma=\rm constant$ or equivalently $g_0N=\rm constant$), there are two degenerate ground states which for $\gamma\ll -1$ describe a Bose-Einstein condensate located in one of the wells of the double-well potential in (\ref{Sll2}) with $s=2$. 

For finite $N$, one can diagonalize the two-mode Hamiltonian (\ref{2modeH}) and the lowest energy eigenstates are not exactly degenerate and they obey the space translation symmetry of the Hamiltonian (\ref{Sll2}) --- for $\gamma \ll -1$ they correspond to
\be
|\tilde \psi_\pm\ra \approx \frac{1}{\sqrt{2}}\left(|N,0\ra\pm|0,N\ra\right),
\label{noon}
\ee
where $N$ bosons occupy either the mode $w_1$ (the state $|N,0\ra$) or the mode $w_2$ (the state $|0,N\ra$). However, the energy splitting between the corresponding eigenenergies, $E_--E_+$, goes exponentially quickly to zero with an increase of $N$ but fixed $\gamma$ \cite{Zin2008,Sacha2015PRA}. Thus, in the mean-field limit if the system is initially prepared in one of the symmetry broken states, $|N,0\ra$ or $|0,N\ra$, it lives in such a state infinitely long. Even if one was able to prepare in the experiment one of the states (\ref{noon}), any decoherence (e.g., measurement of the position of a single boson) would lead to the collapse to one of the symmetry-broken states, $|N,0\ra$ or $|0,N\ra$, because $|\tilde \psi_\pm\ra$ are actually Schr\"odinger cat-like states. The collapse indicates the spontaneous symmetry breaking phenomenon \cite{Sacha2015PRA}.

We would like to stress that not only the two lowest eigenstates of the Hamiltonian (\ref{2modeH}) reveal spontaneous symmetry breaking but all eigenstates up to the so-called symmetry broken edge, $E_{\rm edge}$, which can be estimated by means of the mean-field theory \cite{SachaTC2020} 
\be
E_{\rm edge}-E_\pm=N\frac{J}{2}\left(|\gamma|+\frac{1}{|\gamma|}-2\right)>0.
\ee
This is crucial because in order to call a discrete time crystal a new phase, an extensive number of states of the system should break discrete time translation symmetry.
For $\gamma\ll-1$, the states $|N,0\ra$ and $|0,N\ra$ are good approximation for the degenerate symmetry-broken ground states of (\ref{2modeH}). In the Lipkin-Meshkov-Glick formulation (\ref{LMG}) they correspond to all spins polarized along $z$-axis, i.e., to the eigenstates of $\hat S_z$ with $\pm N/2$ eigenvalues. Low energy excitations are related to transfer of atoms from one potential well to the other, i.e., to eigenstates which can be approximated by $|N-i,i\ra$ and $|i,N-i\ra$ --- in the Lipkin-Meshkov-Glick formulation they are eigenstates of $\hat S_z$ with eigenvalues $\pm(N-i)/2$. In the thermodynamic limit these eigenstates are degenerate up to the energy $E_{\rm edge}$ and reveal spontaneous symmetry breaking.

It is also worth mentioning that low-energy states of the Hamiltonian (\ref{Sll2}) correspond to the $s:1$ resonant driving of the particles in the laboratory frame. That is, low-energy states of  (\ref{Sll2}) are related to $p_i\approx \omega/s$ in the laboratory frame and thus, in the classical description, to the particles which are moving along the ring with a period $s$ times longer than the driving period $T$. For $s=2$ considered in this section this means the $2:1$ resonance.

\subsection{Bethe ansatz approach}
\label{BetheAnsatz}

For $\lambda=0$, eigenstates of (\ref{Sll2}) can be found analytically by means of the Bethe ansatz approach \cite{gaudin_2014}. The basic idea of the approach is the observation that between collisions of the particles, where certain $x_i=x_j$, the particles are described by a product of momentum eigenstates $e^{ik_1x_1}e^{ik_2x_2}\dots e^{ik_Nx_N}$ which has to be symmetrized if the particles are bosons. In order to fulfill boundary conditions imposed by the contact interaction potential and the periodic boundary conditions on a ring, the parameters $k_i$ (called quasi-momenta) of the eigenstates have to satisfy the following Bethe equations
\bea
2\pi k_i+2\sum_{j\ne i}^N\arctan\left(\frac{k_i-k_j}{g_0}\right)=2\pi I_i,
\eea
where the $I_i$'s are arbitrary integers (half integers) for odd (even) $N$ which label the eigenstates of the system.
The sums
\bea
P_{\rm total}&=&\sum_{i=1}^Nk_i,
\\ 
E&=&\sum_{i=1}^N\frac{k_i^2}{2}
\eea
are eigenvalues of the total momentum and energy of the unperturbed LL model, respectively. The corresponding eigenstates read
\be
\tilde \psi_{\vec k}(\vec x)\propto 
\prod_{n<m}\left[
\frac{\partial}{\partial x_n}-\frac{\partial}{\partial x_m}+g_0\;{\rm sgn}(x_n-x_m)
\right]\det[e^{ik_jx_i}].
\label{eigenBethe}
\ee

When the external potential is on, diagonalization of the Hamiltonian (\ref{Sll2}) can be performed in the eigenbasis of the unperturbed LL model (\ref{eigenBethe}). The only non-diagonal matrix elements correspond to the external potential
\be
\lambda\sum_{i=1}^N\la \tilde{\psi}_{\vec k'}| \cos(sx_i)|\tilde\psi_{\vec k}\ra.
\ee
In the remaining part of Sec.~\ref{rotatingLL} we will focus on the case of the double-well potential in (\ref{Sll2}), i.e., $s=2$.

\subsection{Bright soliton regime}
\label{BetheSec}

Let us start with the unperturbed LL model ($\lambda=0$) and apply the mean-field approximation in order to find the ground state of the Hamiltonian (\ref{Sll2}). That is, assume that in the moving frame we look for the lowest energy state in the Hilbert space of product states 
\be
\tilde\psi(\vec x)=\prod_{i=1}^N\phi_0(x_i). 
\label{productsol}
\ee
To find the minimal value of $\la\tilde\psi|\tilde H|\tilde\psi\ra$ we have to minimize the following energy functional 
\be
E=N\int_0^{2\pi}dx\left[\frac12|\partial_x \phi_0(x)|^2+\frac{g_0(N-1)}{2}|\phi_0(x)|^4\right],
\ee
with the constraint $\la\phi_0|\phi_0\ra=1$, which reduces to the solution of the Gross-Pitaevskii equation
\be
-\frac12\partial_x^2\phi_0(x)+g_0(N-1)|\phi_0(x)|^2\phi_0(x)=\mu\phi_0(x),
\ee
where $\mu$ is the chemical potential. For attractively interacting bosons on a ring, there is a critical interaction strength $g_0(N-1)=\pi/2$ when the uniform solution, $\phi_0=1/\sqrt{2\pi}$, becomes unstable and a bright soliton is born. If $2/[g_0(N-1)]\ll 2\pi$, the soliton solution is well approximated by the free space bright soliton wavefunction 
\be
\phi_0(x-q)=\sqrt{\frac{1}{2\xi}}	\cosh^{-1}\left[\frac{x-q}{\xi}\right] ,
\label{phi0sol}
\ee
where  
\be
\xi=\frac{2}{g_0(N-1)},
\ee
and $q$ is the location of the solitonic center which can be arbitrary. The corresponding chemical potential 
\be
\mu=-\frac{g_0^2(N-1)^2}{8}.
\label{mu}
\ee

The Bethe-ansatz approach (see Sec.~\ref{BetheAnsatz}) allows one to obtain the exact many-body ground state which, due to the continuous space translation symmetry of the unperturbed LL Hamiltonian, corresponds to the uniform single-particle probability density $\tilde P(x)$. However, the continuous space translation symmetry is spontaneously broken (e.g., when we measure the particles' positions) and the bright soliton probability profile $|\phi_0(x-q)|^2$ emerges in the density of the detected bosons. The location $q$ of the soliton is randomly determined in the detection process and can be any random position on a ring, i.e., the space translation symmetry is broken spontaneously. 

If the bosons are initially prepared in the bright soliton state, i.e., the Bose-Einstein condensate state (\ref{productsol}) with $\phi_0$ given in (\ref{phi0sol}), the mean-field approach predicts it is a stable stationary state. This prediction is true only in the limit of $N\rightarrow\infty$ and $g_0\rightarrow 0$ with $g_0N=\rm constant$. In the full many-body description, the center of mass of the bosons is described by the {\it free-particle} Hamiltonian, $H_{\rm cm}=P_{\rm cm}^2/(2N)$ \cite{Castin_LesHouches}, and even if initially the center of mass is much better localized than the bright soliton width $\xi$, it starts spreading leading to delocalization of the soliton. However, for a finite but large $N$, the delocalization takes a very long time and has not been observed experimentally yet.

In the presence of the double-well potential  [$\lambda\ne 0$ and $s=2$ in (\ref{Sll2})] we are interested in low-energy eigenstates of the system which can be obtained numerically as described in Sec.~\ref{BetheAnsatz}. In the bright soliton regime, the character of low-energy excitations can be analyzed by employing a quantum description to the center of mass of the soliton. If the double-well potential is weak and the width $\xi$ of the bright soliton smaller than the width $\sigma$ of the single-particle wave-packets $w_1$ or $w_2$ (see Sec.~\ref{2modeSec}), the profile of the bright soliton is not much affected by the presence of the potential but the center of mass behavior is. For $\xi<\sigma\approx 1/\sqrt{\Omega}$, where $\Omega=2\sqrt{\lambda}$ is the frequency of the harmonic approximation of the potential wells, the center of mass of the soliton can be described by the quantum Hamiltonian
\be
H_{\rm cm}=-\frac{1}{2N} \partial_q^2+\lambda N
\int_0^{2\pi} dx |\phi_0(x-q)|^2\cos(2x),
\label{Hcm}
\ee
with $\phi_0$ given in (\ref{phi0sol}). The Hamiltonian (\ref{Hcm}) can be derived by means of  the many-body approach or by an extended Bogoliubov approach where the soliton position is treated in a non-perturbative way \cite{Dziarmaga2004,Weiss2009,Sacha2009}.
The eigenequation $H_{\rm cm}\chi_n(q)=E_n\chi_n(q)$ describes the ground state of the center of mass of the system and its excitations. The latter are the low-energy excitations of the system in the bright soliton regime where bosons form a bound state and taking a single particle from such a bound state costs  energy  given by the modulus of the chemical potential (\ref{mu}) which is larger than the excitations of the center of mass of the system. Thus, in the bright soliton regime the excitations of the system have the character of a {\it single-body} of mass $N$ described by the Hamiltonian (\ref{Hcm}). 

\begin{figure}[t]
\includegraphics[width=0.9\columnwidth]{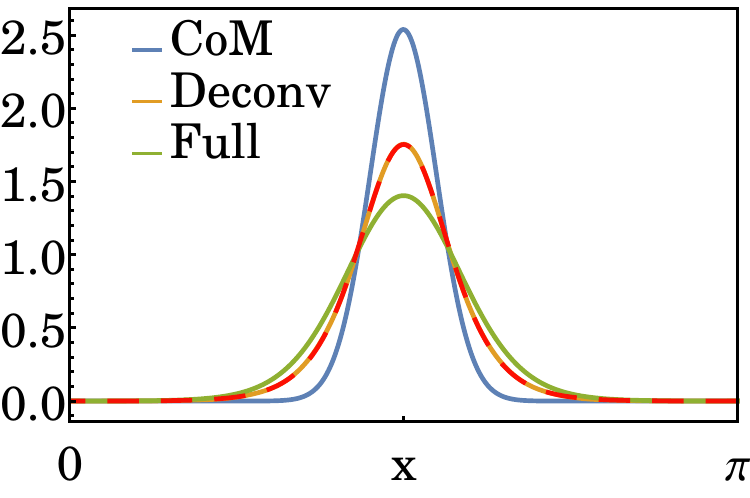}
\caption{The lowest peak is the exact single-particle probability density $\tilde P(x)$ corresponding to one of the symmetry-broken ground states of (\ref{Sll2}) for $N=9$, $\lambda=1.5$ and $g_0(N-1)=-5.33$, cf. Fig.~1 in the Letter. The middle-height curves are the deconvoluted $|\phi_0|^2$ profile (solid line) and the analytical bright soliton profile (dash line) corresponding to $g_0(N-1)=-5.33$. The highest peak is the ground-state probability density, $|\chi_0|^2$, of $H_{\rm cm}$.}
\label{Sfig}
\end{figure}

The strongest interactions analyzed in the Letter are already in the bright soliton regime. However, in order to decribe excitations of  the center of mass of bosons we do not assume that $|\phi_0|^2$ in (\ref{Hcm}) is given by the bright soliton solution (\ref{phi0sol}) but find $|\phi_0|^2$ in a self-consistent way. That is, having the exact single-particle probability density $\tilde P(x)$ correpsonding to one of the symmetry-broken ground states of (\ref{Sll2}) and starting with the ground state $\chi_0(q)$ of the harmonic Hamiltonian, $H_{\rm cm}\approx -\partial_q^2/(2N)+N\Omega^2q^2/2$, we can obtain $|\phi_0|^2$ by solving the equation 
	\be
\tilde P(x)\approx\int_0^{2\pi}dx\;|\phi_0(x-q)|^2|\chi_0(q)|^2, 
\ee 
which is an approximation of the single-particle probability density of the system in the bright soliton state. By means of the Fourier transform we get $|\phi_0|^2$ and substitute it in (\ref{Hcm}) and find a new $\chi_0(q)$ and continue the procedure until $|\phi_0|^2$ converges. A few iterations is sufficient to obtain the final $|\phi_0|^2$ which well reproduces the bright soliton profile (\ref{phi0sol}), see Fig.~\ref{Sfig}. Having $|\phi_0|^2$ we can find the  ground state energy and excited energy levels of (\ref{Hcm}) which describe excitations of the center of mass of the system presented in Fig.~1(a) in the Letter.

\subsection{Example of experimentally attainable parameters}

Ring traps for Bose-Einstein condensates have been realized in various forms, including magnetic ring traps~\cite{Gupta2005}, ring traps involving attractive optical dipole forces using red-detuned light~\cite{bell2016} and ring traps involving repulsive forces using blue-detuned light~\cite{moulder2012,kumar2016}. Optical dipole ring traps have the advantage that they can be used with magnetic Feshbach resonances to tune the s-wave scattering length, which is required for the absolutely stable time crystal. 

In the example calculation in the main text, the parameters are: period of rotating potential $T = \pi/31$ in time units $t_0 = mR^2/\hbar$, tunneling time 302~$T$, and depth of the double-well potential $\lambda = 1.5$ in energy units $E_0 = \hbar^2/mR^2$.  For an optical dipole ring trap with typical radius $R=20~\mu$m \cite{bell2016} containing $~^{39}\text{K}$ atoms, which have a suitable Feshbach resonance at $402~\text{G}$, we have $t_0 = 0.246 \text{s}$ and $E_0/k_B = 0.031~\text{nK}$. Thus, for the above parameters the tunnelling time is 7.5~s, which is shorter than a typical lifetime of a BEC in an optical dipole ring trap, typically about 20-40~s~\cite{bell2016,moulder2012}, the depth of the double-well potential is 0.047~nK, which is readily attainable in optical dipole ring traps, and the frequency of the rotating potential is 40.2~Hz, which is readily attainable in the laboratory.

The choice of double-well potential in the absolutely stable time crystal is flexible provided it is symmetric under $x \rightarrow x + \pi$, i.e., it can be realized in many different ways, for example, by means of a rotating light sheet with the ring cut into two pieces. The frequency of the rotation of the potential can also be chosen almost arbitrarily because after switching to the rotating frame we always end up with the same Hamiltonian (\ref{Sll2}). This means that the resonant momentum $p = \omega/2$ can be chosen as we wish, and we can choose experimentally convenient parameters of Bragg scattering to realize the initial rotation of the atomic cloud. We note that in the case of the kicked Lieb-Liniger model we don't have such flexibility since $\omega$ needs to be sufficiently large to satisfy the rotating wave approximation.
\\~\\
\section{Periodically kicked Lieb-Liniger model}
\label{KICKEDLL}

\subsection{Dicrete time crystals}
\label{kickA}

Let us consider the LL model which is periodically kicked in time,
\be
H=H_{\rm LL}+\lambda T\sum_{i=1}^N \cos(sx_i)\sum_{m=-\infty}^{+\infty}\delta(t-mT),
\label{llk}
\ee
where $T=2\pi/\omega$, and perform the same unitary transformation to the moving frame and the same shift of the momenta as in Sec.~\ref{SecIA} 
\bea
x_i&\rightarrow& x_i+\frac{\omega }{s}t, 
\label{xshift}
\\
p_i&\rightarrow& p_i+\frac{\omega}{s}. 
\label{pshift}
\eea
The resulting Hamiltonian, with a constant term omitted, reads
\bea
\tilde H&=& \sum_{i=1}^{N}\left[ \frac{p_{i}^{2}}{2}+\lambda
\cos(sx_i+\omega t)\sum_{q=-\infty}^{+\infty}e^{iq\omega t}\right] 
\cr &&
+g_{0}\sum_{i<j}^{N}\delta (x_{i}-x_{j}),
\eea
where we have used 
\be
\sum_{m=-\infty}^{+\infty}\delta(t-mT)=\frac{1}{T}\sum_{q=-\infty}^{+\infty}e^{iq\omega t}.
\ee
 Let us switch for a moment to the interaction picture, $e^{i\tilde H_0t}\tilde He^{-i\tilde H_0t}$, where $\tilde H_0=\sum_{i=1}^Np_i^2/2$. Then, the part of the Hamiltonian corresponding to the driving potential reads
\begin{widetext}
\bea
\sum_{i=1}^N\sum_{q=-\infty}^{+\infty}e^{i\tilde H_0t}\lambda \cos(sx_i+\omega t)e^{iq\omega t}e^{-i\tilde H_0t}&=& 
\frac{\lambda}{2}\sum_{i=1}^N\sum_{q=-\infty}^{+\infty}
e^{ip_i^2 t/2}\left(e^{isx_i}e^{i(1+q)\omega t}+e^{-isx_i}e^{-i(1-q)\omega t}\right)e^{-ip_i^2 t/2}.
\cr&&
\label{intpic}
\eea
\end{widetext}
When we calculate matrix elements of (\ref{intpic})
in the momentum basis of bosons, $|\vec k\ra=|k_1,\dots,k_N\ra$, we obtain a sum of the terms in the following form
\be
\frac{\lambda}{2}\left(e^{isx_i}e^{i(1+q)\omega t}+e^{-isx_i}e^{-i(1-q)\omega t}\right)e^{i(k_j^2-k_l^2)t/2}.
\label{intpic1}
\ee
Low-energy states of the Hamiltonian $\tilde H_0$, i.e., states with $k_j\approx 0$, correspond to resonant states in the laboratory frame, i.e., to states with $p_i\approx \omega /s$, cf. (\ref{pshift}). If the periodic driving potential and interactions between bosons are weak, to describe the resonant driving we may restrict to the Hilbert subspace with $k_j^2\ll \omega$. Then, all time-dependent terms (\ref{intpic1}) are quickly oscillating and can be neglected except those with $q=\pm 1$. Such a rotating-wave approximation leads to an effective Hamiltonian of the resonantly kicked LL model which in the Schr\"odinger picture takes the form
\bea
\tilde H\approx 
\sum_{i=1}^{N}\left[ \frac{p_{i}^{2}}{2}+\lambda
\cos(sx_i)\right] +g_{0}\sum_{i<j}^{N}\delta (x_{i}-x_{j}),
\label{kickHRWA}
\eea
which is identical to the Hamiltonian (\ref{Sll2}) of the LL model driven by the rotating lattice potential. Note, however, that in the present case it is an approximate description only and not the exact description as in the case of the rotating potential. The neglected quickly oscillating terms do not allow us to claim that the discrete time crystals in the kicked LL model, which we describe in the Letter, are absolutely stable.  

Another way to obtain the effective Hamiltonian (\ref{kickHRWA}) is to apply the Magnus expansion \cite{Blanes2010}. Kicked models are special as the Magnus expansion can be reduced to the Baker-Campbell-Hausdorff formula. The problem of finding the Floquet Hamiltonian is that of finding a logarithm of the unitary operator of the evolution of the system over a single period $T$. In the moving frame we obtain
\be
\tilde H=\frac{i}{T}\log\left[e^{-i \tilde H_{LL} T}\;e^{-i \lambda T \tilde H_{kick}}\right],
\ee
where 
\be
\tilde H_{kick}=\sum_{i=1}^N \cos(sx_i).
\ee
Employing the Baker-Campbell-Hausdorff formula one gets the following leading terms
\be
\tilde H \approx \tilde H_{LL}+ \lambda \tilde H_{kick} -\frac{i\pi\lambda}{\omega} \;[\tilde H_{LL},\;\tilde H_{kick}],
\ee
where the last term can be omitted if the driving frequency $\omega$ is high, i.e., if the resonant kicking of the bosons corresponds to highly excited momentum states in the laboratory frame, $p_i\approx \omega/s$, cf. (\ref{Eq.TDSEforUtPsi}). This way we arrive at the same effective Hamiltonian as obtained within the rotating-wave approximation, see (\ref{kickHRWA}). We should stress that it is not guaranted that the Magnus expansion converges and therefore there is no guarantee that discrete time crystals in the kicked LL model are absolutely stable.

\subsection{Spontaneous breaking of the space translation symmetry without breaking the time translation symmetry}
\label{SingleKickedB}

If, instead of the transformations (\ref{xshift})-(\ref{pshift}), we apply
\bea
x_i&\rightarrow& x_i+\omega t, 
\label{xshift1}
\\
p_i&\rightarrow& p_i+\omega, 
\label{pshift1}
\eea
then the kicked LL Hamiltonian takes the form 
\bea
\tilde H&=& \sum_{i=1}^{N}\left[ \frac{p_{i}^{2}}{2}+\lambda
\cos(sx_i+s\omega t)\sum_{q=-\infty}^{+\infty}e^{iq\omega t}\right] 
\cr &&
+g_{0}\sum_{i<j}^{N}\delta (x_{i}-x_{j}).
\label{kickfullmov}
\eea
In the interaction picture, the terms analogous to (\ref{intpic1}) read
\be
\frac{\lambda}{2}\left(e^{isx_i}e^{i(s+q)\omega t}+e^{-isx_i}e^{-i(s-q)\omega t}\right)e^{i(k_j^2-k_l^2)t/2},
\ee
and only those with $q=\pm s$ are slowly varying in time and form the effective Hamiltonian of the system within the rotating-wave approximation which is identical to (\ref{kickHRWA}). However, spontaneous breaking of the space translation symmetry in the moving frame does not imply breaking of the time translation symmetry in the laboratory frame because the relation between the single-particle probability density in the laboratory and moving frames is the following, cf. Sec.~\ref{SecIB}
\be
P(x,t)=\tilde P(x-\omega t).
\ee
Thus, in the kicked LL model we can observe a richer interplay between time and space translation symmetry breaking. This is possible because the time-dependent perturbation in (\ref{kickfullmov}) contains many different harmonics in time and many different resonant drivings are possible \cite{Sacha2015PRA,Giergiel2018a}. This is contrary to the time-dependent perturbation in (\ref{Sll1}) where there is only one harmonic in time in the time-dependent perturbation. Resonant driving and formation of resonant islands in classical single-particle phase space, which support discrete time crystal states in the quantum many-body description, are described in the next subsection.

\subsection{Single-particle classical phase space}

\begin{figure}[t]
\includegraphics[width=0.49\columnwidth]{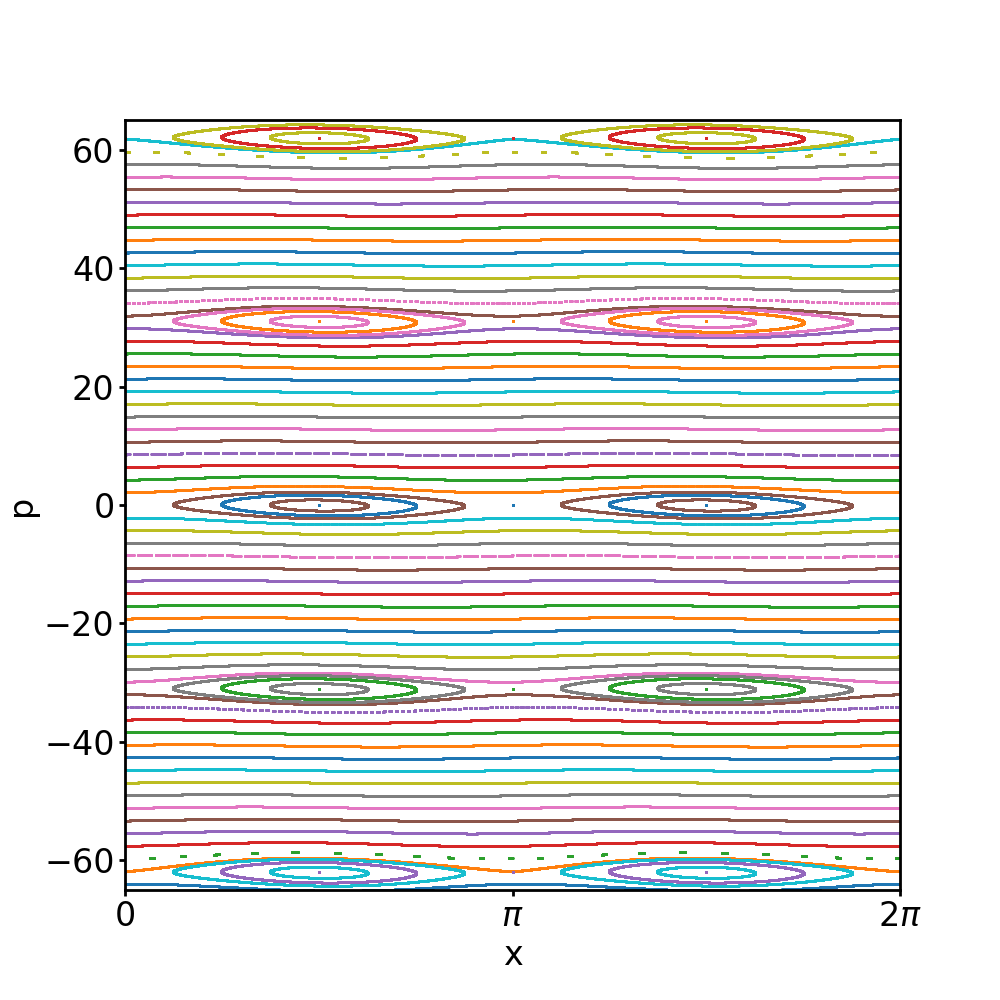}
\includegraphics[width=0.49\columnwidth]{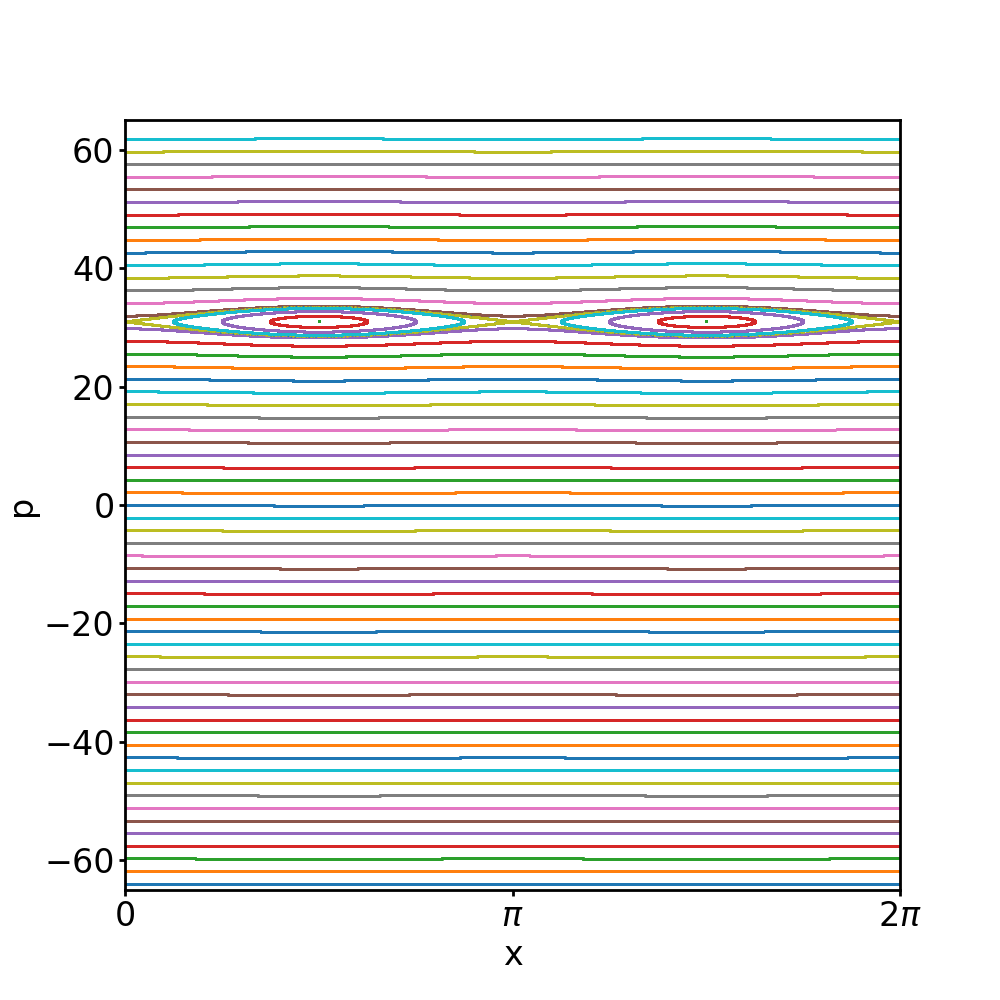}
\includegraphics[width=0.49\columnwidth]{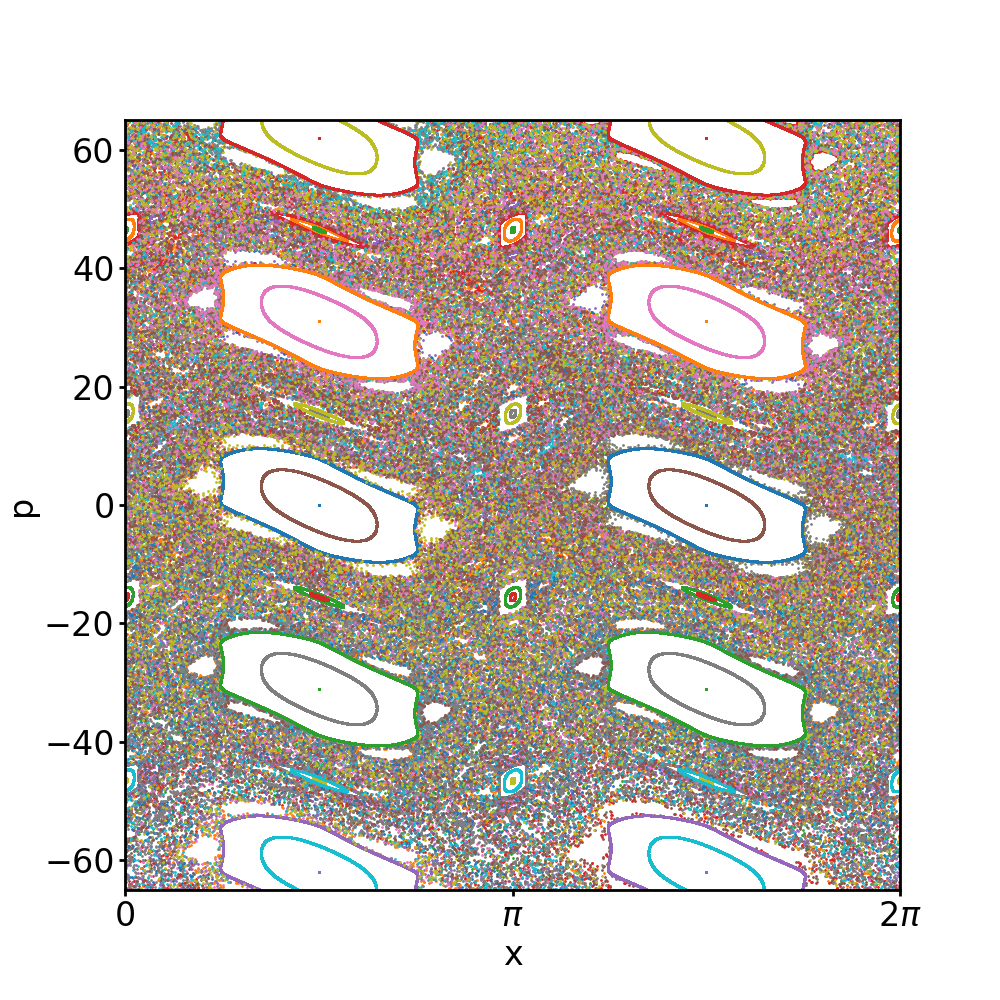}
\includegraphics[width=0.49\columnwidth]{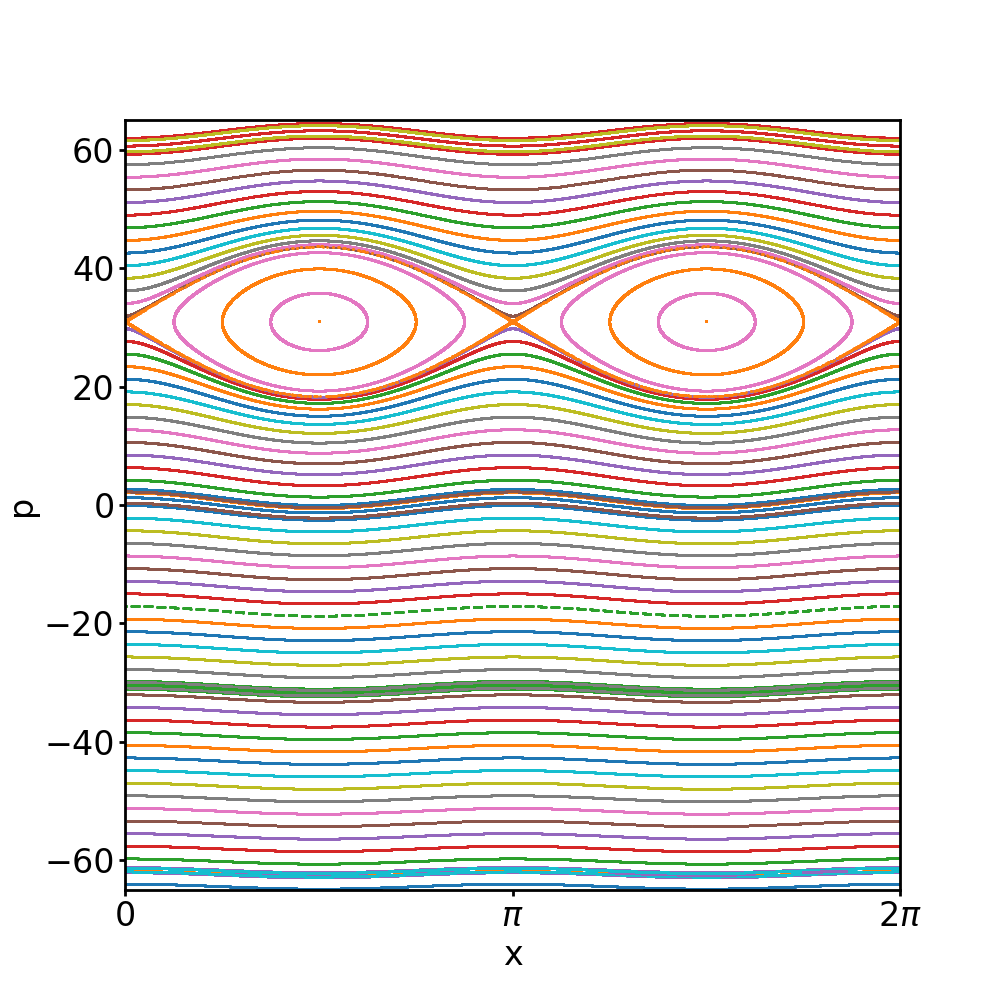}
\caption{Stroboscopic maps (i.e., postion $x$ and momentum $p$ of the particle after each period $T$ of the time evolution of the particle) in the laboratory frame for the kicked single particle (left column) and the single particle driven by the rotating potential (right column) for $s=2$ and $T=\pi/31$. Top row corresponds to $\lambda=1.5$ while bottom row to $\lambda=40$.}
\label{Ssos}
\end{figure}

Let us consider the single-particle counterpart of the kicked LL model
\be
H_{\rm single}=\frac{p^2}{2}+\lambda T\cos(sx)\sum_{m=-\infty}^{+\infty}\delta(t-mT).
\label{SingleKicked}
\ee
The classical motion of the particle can be analyzed with the help of a stroboscopic map where after each evolution period, i.e., at $t_n=nT$, points $(x_n,p_n)$ in classical phase space are plotted. In the kicked system, $(x_n,p_n)$ are solutions of the following Hamilton equations
\bea
x_{n+1}&=&x_n+p_nT, 
\label{sos1}
\\
p_{n+1}&=&p_n+s\lambda T\sin(sx_n).
\label{sos2}
\eea
The particle is moving on the ring, so we restrict to $x_n$ and $x_{n+1}$ modulo $2\pi$. Moreover, if we shift $p_n$ and $p_{n+1}$ by $2\pi/T$, then (\ref{sos2}) does not change but also (\ref{sos1}) does not change because $x_n$ and $x_{n+1}$ are defined modulo $2\pi$. Thus, the phase space structure is periodic along the momentum axis with the period $2\pi/T$. 

In Fig.~\ref{Ssos}(left column) examples of the stroboscopic maps are shown for $s=2$, $T=\pi/31$ and two different values of the strength of the kicking, i.e., $\lambda=1.5$ (used in the Letter) and $\lambda=40$. In the latter case the kicking is so strong that a chaotic sea appears between the clearly visible resonance islands. 

Figure~\ref{Ssos}(right panel) also presents similar stroboscopic maps but for the case when the particle is driven by the rotating lattice potential, i.e., when the single-particle Hamiltonian reads
\be
H_{\rm single}=\frac{p^2}{2}+\lambda \cos(2x-\omega t).
\label{SingleAbsol}
\ee
Then, there are only two islands corresponding to the $2:1$ resonant driving of the particle, i.e., a particle starting at the center of one of the islands at $t=0$ jumps to the other island at $t=T$ and returns to the initial island at $t=2T$. In the quantum description, if the particle is initially prepared in a wave-packet localized in one of the islands, it jumps between the islands every period $T$ for some time but such a $2T$-periodic evolution is broken due to the tunneling process. Only if we switch to the many-body case and the particles sufficiently strongly attract each other, the tunneling process stops and a discrete time crystal is formed \cite{Sacha2015PRA}.

In the case of the rotating potential (\ref{SingleAbsol}), there are only $2:1$ resonance islands and regardless of how strong the driving is, we do not see irregular dynamics. Switching to the moving frame, $x\rightarrow x+\omega t/2$, and performing the shift in the momentum, $p\rightarrow p+\omega/2$, we obtain the time-independent Hamiltonian of the particle in the double well potential. Low-energy eigenstates of the quantum many-body counterpart of such a system correspond to the absolutely stable discrete time crystals which we consider in the Letter.

In the case of the kicked particle (\ref{SingleKicked}), we can see a few different resonances in Fig.~\ref{Ssos}, i.e., co-rotating ($p=\pi/T$) and counter-rotating ($p=-\pi/T$) $2:1$ resonances and co-rotating ($p=2\pi/T$) and counter-rotating ($p=-2\pi/T$) $1:1$ resonances. The $1:1$ resonances allow us to realize spontaneous breaking of the discrete space translation symmetry (i.e. the symmetry under the shift $x\rightarrow x+\pi$) without breaking the discrete time translation symmetry if the many-body system is appropriately driven as described in Sec.~\ref{SingleKickedB}. The presence of different resonances offers richer interplay between time and space symmetry breaking but on the other hand it does not allow one to claim that discrete time crystals in the kicked LL model are absolutely stable. Indeed, apart from a resonant term there are always non-resonant ones which cannot be eliminated and can lead to decay of the crystal on a long time scale.


\begin{thebibliography}{65}
\expandafter\ifx\csname natexlab\endcsname\relax\def\natexlab#1{#1}\fi
\expandafter\ifx\csname bibnamefont\endcsname\relax
  \def\bibnamefont#1{#1}\fi
\expandafter\ifx\csname bibfnamefont\endcsname\relax
  \def\bibfnamefont#1{#1}\fi
\expandafter\ifx\csname citenamefont\endcsname\relax
  \def\citenamefont#1{#1}\fi
\expandafter\ifx\csname url\endcsname\relax
  \def\url#1{\texttt{#1}}\fi
\expandafter\ifx\csname urlprefix\endcsname\relax\def\urlprefix{URL }\fi
\providecommand{\bibinfo}[2]{#2}
\providecommand{\eprint}[2][]{\url{#2}}

\bibitem[{\citenamefont{Wilczek}(2012)}]{WilczekPRL2012}
\bibinfo{author}{\bibfnamefont{F.}~\bibnamefont{Wilczek}},
  \bibinfo{journal}{Phys. Rev. Lett.} \textbf{\bibinfo{volume}{109}},
  \bibinfo{pages}{160401} (\bibinfo{year}{2012}).

\bibitem[{\citenamefont{Sacha}(2015{\natexlab{a}})}]{Sacha2015PRA}
\bibinfo{author}{\bibfnamefont{K.}~\bibnamefont{Sacha}},
  \bibinfo{journal}{Phys. Rev. A} \textbf{\bibinfo{volume}{91}},
  \bibinfo{pages}{033617} (\bibinfo{year}{2015}{\natexlab{a}}),
  \urlprefix\url{http://link.aps.org/doi/10.1103/PhysRevA.91.033617}.

\bibitem[{\citenamefont{Khemani et~al.}(2016)\citenamefont{Khemani, Lazarides,
  Moessner, and Sondhi}}]{KhemaniPRL2016}
\bibinfo{author}{\bibfnamefont{V.}~\bibnamefont{Khemani}},
  \bibinfo{author}{\bibfnamefont{A.}~\bibnamefont{Lazarides}},
  \bibinfo{author}{\bibfnamefont{R.}~\bibnamefont{Moessner}}, \bibnamefont{and}
  \bibinfo{author}{\bibfnamefont{S.~L.} \bibnamefont{Sondhi}},
  \bibinfo{journal}{Phys. Rev. Lett.} \textbf{\bibinfo{volume}{116}},
  \bibinfo{pages}{250401} (\bibinfo{year}{2016}).

\bibitem[{\citenamefont{Else et~al.}(2016)\citenamefont{Else, Bauer, and
  Nayak}}]{ElsePRL2016}
\bibinfo{author}{\bibfnamefont{D.~V.} \bibnamefont{Else}},
  \bibinfo{author}{\bibfnamefont{B.}~\bibnamefont{Bauer}}, \bibnamefont{and}
  \bibinfo{author}{\bibfnamefont{C.}~\bibnamefont{Nayak}},
  \bibinfo{journal}{Phys. Rev. Lett.} \textbf{\bibinfo{volume}{117}},
  \bibinfo{pages}{090402} (\bibinfo{year}{2016}).

\bibitem[{\citenamefont{Yao et~al.}(2017)\citenamefont{Yao, Potter, Potirniche,
  and Vishwanath}}]{YaoPRL2017}
\bibinfo{author}{\bibfnamefont{N.~Y.} \bibnamefont{Yao}},
  \bibinfo{author}{\bibfnamefont{A.~C.} \bibnamefont{Potter}},
  \bibinfo{author}{\bibfnamefont{I.-D.} \bibnamefont{Potirniche}},
  \bibnamefont{and}
  \bibinfo{author}{\bibfnamefont{A.}~\bibnamefont{Vishwanath}},
  \bibinfo{journal}{Phys. Rev. Lett.} \textbf{\bibinfo{volume}{118}},
  \bibinfo{pages}{030401} (\bibinfo{year}{2017}).

\bibitem[{\citenamefont{Sacha}(2020)}]{SachaTC2020}
\bibinfo{author}{\bibfnamefont{K.}~\bibnamefont{Sacha}},
  \emph{\bibinfo{title}{Time Crystals}} (\bibinfo{publisher}{Springer
  International Publishing}, \bibinfo{address}{Switzerland, Cham},
  \bibinfo{year}{2020}), ISBN \bibinfo{isbn}{978-3-030-52523-1},
  \urlprefix\url{https://doi.org/10.1007/978-3-030-52523-1}.

\bibitem[{\citenamefont{D'Alessio and Rigol}(2014)}]{DAlessioPRX2014}
\bibinfo{author}{\bibfnamefont{L.}~\bibnamefont{D'Alessio}} \bibnamefont{and}
  \bibinfo{author}{\bibfnamefont{M.}~\bibnamefont{Rigol}},
  \bibinfo{journal}{Phys. Rev. X} \textbf{\bibinfo{volume}{4}},
  \bibinfo{pages}{041048} (\bibinfo{year}{2014}).

\bibitem[{\citenamefont{Lazarides et~al.}(2014)\citenamefont{Lazarides, Das,
  and Moessner}}]{LazaridesPRE2014}
\bibinfo{author}{\bibfnamefont{A.}~\bibnamefont{Lazarides}},
  \bibinfo{author}{\bibfnamefont{A.}~\bibnamefont{Das}}, \bibnamefont{and}
  \bibinfo{author}{\bibfnamefont{R.}~\bibnamefont{Moessner}},
  \bibinfo{journal}{Phys. Rev. E} \textbf{\bibinfo{volume}{90}},
  \bibinfo{pages}{012110} (\bibinfo{year}{2014}).

\bibitem[{\citenamefont{Ponte et~al.}(2015{\natexlab{a}})\citenamefont{Ponte,
  Chandran, Papi{\'c}, and Abanin}}]{PonteAPA2015}
\bibinfo{author}{\bibfnamefont{P.}~\bibnamefont{Ponte}},
  \bibinfo{author}{\bibfnamefont{A.}~\bibnamefont{Chandran}},
  \bibinfo{author}{\bibfnamefont{Z.}~\bibnamefont{Papi{\'c}}},
  \bibnamefont{and} \bibinfo{author}{\bibfnamefont{D.~A.}
  \bibnamefont{Abanin}}, \bibinfo{journal}{Ann. Phys. (Amsterdam)}
  \textbf{\bibinfo{volume}{353}}, \bibinfo{pages}{196}
  (\bibinfo{year}{2015}{\natexlab{a}}).

\bibitem[{\citenamefont{Rigol et~al.}(2007)\citenamefont{Rigol, Dunjko,
  Yurovsky, and Olshanii}}]{RigolPRL2007}
\bibinfo{author}{\bibfnamefont{M.}~\bibnamefont{Rigol}},
  \bibinfo{author}{\bibfnamefont{V.}~\bibnamefont{Dunjko}},
  \bibinfo{author}{\bibfnamefont{V.}~\bibnamefont{Yurovsky}}, \bibnamefont{and}
  \bibinfo{author}{\bibfnamefont{M.}~\bibnamefont{Olshanii}},
  \bibinfo{journal}{Phys. Rev. Lett.} \textbf{\bibinfo{volume}{98}},
  \bibinfo{pages}{050405} (\bibinfo{year}{2007}).

\bibitem[{\citenamefont{Santos and Rigol}(2010{\natexlab{a}})}]{SantosPRE2010a}
\bibinfo{author}{\bibfnamefont{L.~F.} \bibnamefont{Santos}} \bibnamefont{and}
  \bibinfo{author}{\bibfnamefont{M.}~\bibnamefont{Rigol}},
  \bibinfo{journal}{Phys. Rev. E} \textbf{\bibinfo{volume}{81}},
  \bibinfo{pages}{036206} (\bibinfo{year}{2010}{\natexlab{a}}).

\bibitem[{\citenamefont{Rigol and Santos}(2010)}]{MarcosPRA2010}
\bibinfo{author}{\bibfnamefont{M.}~\bibnamefont{Rigol}} \bibnamefont{and}
  \bibinfo{author}{\bibfnamefont{L.~F.} \bibnamefont{Santos}},
  \bibinfo{journal}{Phys. Rev. A} \textbf{\bibinfo{volume}{82}},
  \bibinfo{pages}{011604} (\bibinfo{year}{2010}).

\bibitem[{\citenamefont{Santos and Rigol}(2010{\natexlab{b}})}]{SantosPRE2010b}
\bibinfo{author}{\bibfnamefont{L.~F.} \bibnamefont{Santos}} \bibnamefont{and}
  \bibinfo{author}{\bibfnamefont{M.}~\bibnamefont{Rigol}},
  \bibinfo{journal}{Phys. Rev. E} \textbf{\bibinfo{volume}{82}},
  \bibinfo{pages}{031130} (\bibinfo{year}{2010}{\natexlab{b}}).

\bibitem[{\citenamefont{Cassidy et~al.}(2011)\citenamefont{Cassidy, Clark, and
  Rigol}}]{CassidyPRL2011}
\bibinfo{author}{\bibfnamefont{A.~C.} \bibnamefont{Cassidy}},
  \bibinfo{author}{\bibfnamefont{C.~W.} \bibnamefont{Clark}}, \bibnamefont{and}
  \bibinfo{author}{\bibfnamefont{M.}~\bibnamefont{Rigol}},
  \bibinfo{journal}{Phys. Rev. Lett.} \textbf{\bibinfo{volume}{106}},
  \bibinfo{pages}{140405} (\bibinfo{year}{2011}).

\bibitem[{\citenamefont{Caux and Essler}(2013)}]{CauxPRL2013}
\bibinfo{author}{\bibfnamefont{J.-S.} \bibnamefont{Caux}} \bibnamefont{and}
  \bibinfo{author}{\bibfnamefont{F.~H.~L.} \bibnamefont{Essler}},
  \bibinfo{journal}{Phys. Rev. Lett.} \textbf{\bibinfo{volume}{110}},
  \bibinfo{pages}{257203} (\bibinfo{year}{2013}).

\bibitem[{\citenamefont{Nandkishore and Huse}(2015)}]{Nandkishore2015}
\bibinfo{author}{\bibfnamefont{R.}~\bibnamefont{Nandkishore}} \bibnamefont{and}
  \bibinfo{author}{\bibfnamefont{D.~A.} \bibnamefont{Huse}},
  \bibinfo{journal}{Annu. Rev. Condens. Matter Phys.}
  \textbf{\bibinfo{volume}{6}}, \bibinfo{pages}{15} (\bibinfo{year}{2015}).

\bibitem[{\citenamefont{Altman and Vosk}(2015)}]{Altman2015}
\bibinfo{author}{\bibfnamefont{E.}~\bibnamefont{Altman}} \bibnamefont{and}
  \bibinfo{author}{\bibfnamefont{R.}~\bibnamefont{Vosk}},
  \bibinfo{journal}{Annu. Rev. Condens. Matter Phys.}
  \textbf{\bibinfo{volume}{6}}, \bibinfo{pages}{383} (\bibinfo{year}{2015}).

\bibitem[{\citenamefont{D'Alessio et~al.}(2016)\citenamefont{D'Alessio, Kafri,
  Polkovnikov, and Rigol}}]{DAlessio2016}
\bibinfo{author}{\bibfnamefont{L.}~\bibnamefont{D'Alessio}},
  \bibinfo{author}{\bibfnamefont{Y.}~\bibnamefont{Kafri}},
  \bibinfo{author}{\bibfnamefont{A.}~\bibnamefont{Polkovnikov}},
  \bibnamefont{and} \bibinfo{author}{\bibfnamefont{M.}~\bibnamefont{Rigol}},
  \bibinfo{journal}{Adv. Phys.} \textbf{\bibinfo{volume}{65}},
  \bibinfo{pages}{239} (\bibinfo{year}{2016}).

\bibitem[{\citenamefont{Mierzejewski et~al.}(2018)\citenamefont{Mierzejewski,
  Kozarzewski, and Prelov\ifmmode~\check{s}\else
  \v{s}\fi{}ek}}]{Mierzejewski2018}
\bibinfo{author}{\bibfnamefont{M.}~\bibnamefont{Mierzejewski}},
  \bibinfo{author}{\bibfnamefont{M.}~\bibnamefont{Kozarzewski}},
  \bibnamefont{and}
  \bibinfo{author}{\bibfnamefont{P.}~\bibnamefont{Prelov\ifmmode~\check{s}\else
  \v{s}\fi{}ek}}, \bibinfo{journal}{Phys. Rev. B}
  \textbf{\bibinfo{volume}{97}}, \bibinfo{pages}{064204}
  (\bibinfo{year}{2018}),
  \urlprefix\url{https://link.aps.org/doi/10.1103/PhysRevB.97.064204}.

\bibitem[{\citenamefont{Else et~al.}(2017)\citenamefont{Else, Bauer, and
  Nayak}}]{ElsePRX2017}
\bibinfo{author}{\bibfnamefont{D.~V.} \bibnamefont{Else}},
  \bibinfo{author}{\bibfnamefont{B.}~\bibnamefont{Bauer}}, \bibnamefont{and}
  \bibinfo{author}{\bibfnamefont{C.}~\bibnamefont{Nayak}},
  \bibinfo{journal}{Phys. Rev. X} \textbf{\bibinfo{volume}{7}},
  \bibinfo{pages}{011026} (\bibinfo{year}{2017}).

\bibitem[{\citenamefont{Ponte et~al.}(2015{\natexlab{b}})\citenamefont{Ponte,
  Papi\ifmmode~\acute{c}\else \'{c}\fi{}, Huveneers, and
  Abanin}}]{PontePRL2015}
\bibinfo{author}{\bibfnamefont{P.}~\bibnamefont{Ponte}},
  \bibinfo{author}{\bibfnamefont{Z.}~\bibnamefont{Papi\ifmmode~\acute{c}\else
  \'{c}\fi{}}}, \bibinfo{author}{\bibfnamefont{F.~M.~C.}
  \bibnamefont{Huveneers}}, \bibnamefont{and}
  \bibinfo{author}{\bibfnamefont{D.~A.} \bibnamefont{Abanin}},
  \bibinfo{journal}{Phys. Rev. Lett.} \textbf{\bibinfo{volume}{114}},
  \bibinfo{pages}{140401} (\bibinfo{year}{2015}{\natexlab{b}}).

\bibitem[{\citenamefont{Lazarides et~al.}(2015)\citenamefont{Lazarides, Das,
  and Moessner}}]{LazaridesPRL2015}
\bibinfo{author}{\bibfnamefont{A.}~\bibnamefont{Lazarides}},
  \bibinfo{author}{\bibfnamefont{A.}~\bibnamefont{Das}}, \bibnamefont{and}
  \bibinfo{author}{\bibfnamefont{R.}~\bibnamefont{Moessner}},
  \bibinfo{journal}{Phys. Rev. Lett.} \textbf{\bibinfo{volume}{115}},
  \bibinfo{pages}{030402} (\bibinfo{year}{2015}).

\bibitem[{\citenamefont{Abanin et~al.}(2016)\citenamefont{Abanin, Roeck, and
  Huveneers}}]{AbaninAP2016}
\bibinfo{author}{\bibfnamefont{D.}~\bibnamefont{Abanin}},
  \bibinfo{author}{\bibfnamefont{W.~D.} \bibnamefont{Roeck}}, \bibnamefont{and}
  \bibinfo{author}{\bibfnamefont{F.}~\bibnamefont{Huveneers}},
  \bibinfo{journal}{Ann. Phys.} \textbf{\bibinfo{volume}{372}},
  \bibinfo{pages}{1} (\bibinfo{year}{2016}).

\bibitem[{\citenamefont{Sierant et~al.}(2023)\citenamefont{Sierant, Lewenstein,
  Scardicchio, and Zakrzewski}}]{Sierant2023}
\bibinfo{author}{\bibfnamefont{P.}~\bibnamefont{Sierant}},
  \bibinfo{author}{\bibfnamefont{M.}~\bibnamefont{Lewenstein}},
  \bibinfo{author}{\bibfnamefont{A.}~\bibnamefont{Scardicchio}},
  \bibnamefont{and}
  \bibinfo{author}{\bibfnamefont{J.}~\bibnamefont{Zakrzewski}},
  \bibinfo{journal}{Phys. Rev. B} \textbf{\bibinfo{volume}{107}},
  \bibinfo{pages}{115132} (\bibinfo{year}{2023}),
  \urlprefix\url{https://link.aps.org/doi/10.1103/PhysRevB.107.115132}.

\bibitem[{\citenamefont{Zhang et~al.}(2017)\citenamefont{Zhang, Hess,
  Kyprianidis, Becker, Lee, Smith, Pagano, Potirniche, Potter, Vishwanath
  et~al.}}]{ZhangNature2017}
\bibinfo{author}{\bibfnamefont{J.}~\bibnamefont{Zhang}},
  \bibinfo{author}{\bibfnamefont{P.~W.} \bibnamefont{Hess}},
  \bibinfo{author}{\bibfnamefont{A.}~\bibnamefont{Kyprianidis}},
  \bibinfo{author}{\bibfnamefont{P.}~\bibnamefont{Becker}},
  \bibinfo{author}{\bibfnamefont{A.}~\bibnamefont{Lee}},
  \bibinfo{author}{\bibfnamefont{J.}~\bibnamefont{Smith}},
  \bibinfo{author}{\bibfnamefont{G.}~\bibnamefont{Pagano}},
  \bibinfo{author}{\bibfnamefont{I.-D.} \bibnamefont{Potirniche}},
  \bibinfo{author}{\bibfnamefont{A.~C.} \bibnamefont{Potter}},
  \bibinfo{author}{\bibfnamefont{A.}~\bibnamefont{Vishwanath}},
  \bibnamefont{et~al.}, \bibinfo{journal}{Nature}
  \textbf{\bibinfo{volume}{543}}, \bibinfo{pages}{217} (\bibinfo{year}{2017}).

\bibitem[{\citenamefont{Choi et~al.}(2017)\citenamefont{Choi, Choi, Landig,
  Kucsko, Zhou, Isoya, Jelezko, Onoda, Sumiya, Khemani
  et~al.}}]{ChoiNature2017}
\bibinfo{author}{\bibfnamefont{S.}~\bibnamefont{Choi}},
  \bibinfo{author}{\bibfnamefont{J.}~\bibnamefont{Choi}},
  \bibinfo{author}{\bibfnamefont{R.}~\bibnamefont{Landig}},
  \bibinfo{author}{\bibfnamefont{G.}~\bibnamefont{Kucsko}},
  \bibinfo{author}{\bibfnamefont{H.}~\bibnamefont{Zhou}},
  \bibinfo{author}{\bibfnamefont{J.}~\bibnamefont{Isoya}},
  \bibinfo{author}{\bibfnamefont{F.}~\bibnamefont{Jelezko}},
  \bibinfo{author}{\bibfnamefont{S.}~\bibnamefont{Onoda}},
  \bibinfo{author}{\bibfnamefont{H.}~\bibnamefont{Sumiya}},
  \bibinfo{author}{\bibfnamefont{V.}~\bibnamefont{Khemani}},
  \bibnamefont{et~al.}, \bibinfo{journal}{Nature}
  \textbf{\bibinfo{volume}{543}}, \bibinfo{pages}{221} (\bibinfo{year}{2017}).

\bibitem[{\citenamefont{Pal et~al.}(2018)\citenamefont{Pal, Nishad, Mahesh, and
  Sreejith}}]{Pal2018}
\bibinfo{author}{\bibfnamefont{S.}~\bibnamefont{Pal}},
  \bibinfo{author}{\bibfnamefont{N.}~\bibnamefont{Nishad}},
  \bibinfo{author}{\bibfnamefont{T.~S.} \bibnamefont{Mahesh}},
  \bibnamefont{and} \bibinfo{author}{\bibfnamefont{G.~J.}
  \bibnamefont{Sreejith}}, \bibinfo{journal}{Phys. Rev. Lett.}
  \textbf{\bibinfo{volume}{120}}, \bibinfo{pages}{180602}
  (\bibinfo{year}{2018}),
  \urlprefix\url{https://link.aps.org/doi/10.1103/PhysRevLett.120.180602}.

\bibitem[{\citenamefont{Rovny et~al.}(2018)\citenamefont{Rovny, Blum, and
  Barrett}}]{Rovny2018}
\bibinfo{author}{\bibfnamefont{J.}~\bibnamefont{Rovny}},
  \bibinfo{author}{\bibfnamefont{R.~L.} \bibnamefont{Blum}}, \bibnamefont{and}
  \bibinfo{author}{\bibfnamefont{S.~E.} \bibnamefont{Barrett}},
  \bibinfo{journal}{Phys. Rev. Lett.} \textbf{\bibinfo{volume}{120}},
  \bibinfo{pages}{180603} (\bibinfo{year}{2018}),
  \urlprefix\url{https://link.aps.org/doi/10.1103/PhysRevLett.120.180603}.

\bibitem[{\citenamefont{Smits et~al.}(2018)\citenamefont{Smits, Liao, Stoof,
  and van~der Straten}}]{Smits2018}
\bibinfo{author}{\bibfnamefont{J.}~\bibnamefont{Smits}},
  \bibinfo{author}{\bibfnamefont{L.}~\bibnamefont{Liao}},
  \bibinfo{author}{\bibfnamefont{H.~T.~C.} \bibnamefont{Stoof}},
  \bibnamefont{and} \bibinfo{author}{\bibfnamefont{P.}~\bibnamefont{van~der
  Straten}}, \bibinfo{journal}{Phys. Rev. Lett.}
  \textbf{\bibinfo{volume}{121}}, \bibinfo{pages}{185301}
  (\bibinfo{year}{2018}),
  \urlprefix\url{https://link.aps.org/doi/10.1103/PhysRevLett.121.185301}.

\bibitem[{\citenamefont{Mi et~al.}(2022)\citenamefont{Mi, Ippoliti, Quintana,
  Greene, Chen, Gross, Arute, Arya, Atalaya, Babbush et~al.}}]{Mi2022}
\bibinfo{author}{\bibfnamefont{X.}~\bibnamefont{Mi}},
  \bibinfo{author}{\bibfnamefont{M.}~\bibnamefont{Ippoliti}},
  \bibinfo{author}{\bibfnamefont{C.}~\bibnamefont{Quintana}},
  \bibinfo{author}{\bibfnamefont{A.}~\bibnamefont{Greene}},
  \bibinfo{author}{\bibfnamefont{Z.}~\bibnamefont{Chen}},
  \bibinfo{author}{\bibfnamefont{J.}~\bibnamefont{Gross}},
  \bibinfo{author}{\bibfnamefont{F.}~\bibnamefont{Arute}},
  \bibinfo{author}{\bibfnamefont{K.}~\bibnamefont{Arya}},
  \bibinfo{author}{\bibfnamefont{J.}~\bibnamefont{Atalaya}},
  \bibinfo{author}{\bibfnamefont{R.}~\bibnamefont{Babbush}},
  \bibnamefont{et~al.}, \bibinfo{journal}{Nature}
  \textbf{\bibinfo{volume}{601}}, \bibinfo{pages}{531} (\bibinfo{year}{2022}),
  ISSN \bibinfo{issn}{1476-4687},
  \urlprefix\url{https://doi.org/10.1038/s41586-021-04257-w}.

\bibitem[{\citenamefont{{Randall} et~al.}(2021)\citenamefont{{Randall},
  {Bradley}, {van der Gronden}, {Galicia}, {Abobeih}, {Markham}, {Twitchen},
  {Machado}, {Yao}, and {Taminiau}}}]{Randall2021}
\bibinfo{author}{\bibfnamefont{J.}~\bibnamefont{{Randall}}},
  \bibinfo{author}{\bibfnamefont{C.~E.} \bibnamefont{{Bradley}}},
  \bibinfo{author}{\bibfnamefont{F.~V.} \bibnamefont{{van der Gronden}}},
  \bibinfo{author}{\bibfnamefont{A.}~\bibnamefont{{Galicia}}},
  \bibinfo{author}{\bibfnamefont{M.~H.} \bibnamefont{{Abobeih}}},
  \bibinfo{author}{\bibfnamefont{M.}~\bibnamefont{{Markham}}},
  \bibinfo{author}{\bibfnamefont{D.~J.} \bibnamefont{{Twitchen}}},
  \bibinfo{author}{\bibfnamefont{F.}~\bibnamefont{{Machado}}},
  \bibinfo{author}{\bibfnamefont{N.~Y.} \bibnamefont{{Yao}}}, \bibnamefont{and}
  \bibinfo{author}{\bibfnamefont{T.~H.} \bibnamefont{{Taminiau}}},
  \bibinfo{journal}{Science} \textbf{\bibinfo{volume}{374}},
  \bibinfo{pages}{1474} (\bibinfo{year}{2021}), \eprint{2107.00736}.

\bibitem[{\citenamefont{Frey and Rachel}(2022)}]{Frey2022}
\bibinfo{author}{\bibfnamefont{P.}~\bibnamefont{Frey}} \bibnamefont{and}
  \bibinfo{author}{\bibfnamefont{S.}~\bibnamefont{Rachel}},
  \bibinfo{journal}{Science Advances} \textbf{\bibinfo{volume}{8}},
  \bibinfo{pages}{eabm7652} (\bibinfo{year}{2022}),
  \eprint{https://www.science.org/doi/pdf/10.1126/sciadv.abm7652},
  \urlprefix\url{https://www.science.org/doi/abs/10.1126/sciadv.abm7652}.

\bibitem[{\citenamefont{{Imbrie}}(2016)}]{Imbrie2016}
\bibinfo{author}{\bibfnamefont{J.~Z.} \bibnamefont{{Imbrie}}},
  \bibinfo{journal}{Journal of Statistical Physics}
  \textbf{\bibinfo{volume}{163}}, \bibinfo{pages}{998} (\bibinfo{year}{2016}),
  \eprint{1403.7837}.

\bibitem[{\citenamefont{Sels and Polkovnikov}(2021)}]{Sels2021}
\bibinfo{author}{\bibfnamefont{D.}~\bibnamefont{Sels}} \bibnamefont{and}
  \bibinfo{author}{\bibfnamefont{A.}~\bibnamefont{Polkovnikov}},
  \bibinfo{journal}{Phys. Rev. E} \textbf{\bibinfo{volume}{104}},
  \bibinfo{pages}{054105} (\bibinfo{year}{2021}),
  \urlprefix\url{https://link.aps.org/doi/10.1103/PhysRevE.104.054105}.

\bibitem[{\citenamefont{Wang et~al.}(2021{\natexlab{a}})\citenamefont{Wang,
  Hannaford, and Dalton}}]{WangNJP2021}
\bibinfo{author}{\bibfnamefont{J.}~\bibnamefont{Wang}},
  \bibinfo{author}{\bibfnamefont{P.}~\bibnamefont{Hannaford}},
  \bibnamefont{and} \bibinfo{author}{\bibfnamefont{B.~J.}
  \bibnamefont{Dalton}}, \bibinfo{journal}{New J. Phys.}
  \textbf{\bibinfo{volume}{23}}, \bibinfo{pages}{063012}
  (\bibinfo{year}{2021}{\natexlab{a}}).

\bibitem[{\citenamefont{Kuro{\'{s}} et~al.}(2020)\citenamefont{Kuro{\'{s}},
  Mukherjee, Golletz, Sauvage, Giergiel, Mintert, and Sacha}}]{Kuros2020}
\bibinfo{author}{\bibfnamefont{A.}~\bibnamefont{Kuro{\'{s}}}},
  \bibinfo{author}{\bibfnamefont{R.}~\bibnamefont{Mukherjee}},
  \bibinfo{author}{\bibfnamefont{W.}~\bibnamefont{Golletz}},
  \bibinfo{author}{\bibfnamefont{F.}~\bibnamefont{Sauvage}},
  \bibinfo{author}{\bibfnamefont{K.}~\bibnamefont{Giergiel}},
  \bibinfo{author}{\bibfnamefont{F.}~\bibnamefont{Mintert}}, \bibnamefont{and}
  \bibinfo{author}{\bibfnamefont{K.}~\bibnamefont{Sacha}},
  \bibinfo{journal}{New Journal of Physics} \textbf{\bibinfo{volume}{22}},
  \bibinfo{pages}{095001} (\bibinfo{year}{2020}),
  \urlprefix\url{https://doi.org/10.1088/1367-2630/abb03e}.

\bibitem[{\citenamefont{Wang et~al.}(2021{\natexlab{b}})\citenamefont{Wang,
  Sacha, Hannaford, and Dalton}}]{WangPRA2021}
\bibinfo{author}{\bibfnamefont{J.}~\bibnamefont{Wang}},
  \bibinfo{author}{\bibfnamefont{K.}~\bibnamefont{Sacha}},
  \bibinfo{author}{\bibfnamefont{P.}~\bibnamefont{Hannaford}},
  \bibnamefont{and} \bibinfo{author}{\bibfnamefont{B.~J.}
  \bibnamefont{Dalton}}, \bibinfo{journal}{Phys. Rev. A}
  \textbf{\bibinfo{volume}{104}}, \bibinfo{pages}{053327}
  (\bibinfo{year}{2021}{\natexlab{b}}),
  \urlprefix\url{https://link.aps.org/doi/10.1103/PhysRevA.104.053327}.

\bibitem[{SM()}]{SM}
\bibinfo{note}{See Supplemental Material}.

\bibitem[{\citenamefont{Gupta et~al.}(2005)\citenamefont{Gupta, Murch, Moore,
  Purdy, and Stamper-Kurn}}]{Gupta2005}
\bibinfo{author}{\bibfnamefont{S.}~\bibnamefont{Gupta}},
  \bibinfo{author}{\bibfnamefont{K.~W.} \bibnamefont{Murch}},
  \bibinfo{author}{\bibfnamefont{K.~L.} \bibnamefont{Moore}},
  \bibinfo{author}{\bibfnamefont{T.~P.} \bibnamefont{Purdy}}, \bibnamefont{and}
  \bibinfo{author}{\bibfnamefont{D.~M.} \bibnamefont{Stamper-Kurn}},
  \bibinfo{journal}{Phys. Rev. Lett.} \textbf{\bibinfo{volume}{95}},
  \bibinfo{pages}{143201} (\bibinfo{year}{2005}),
  \urlprefix\url{https://link.aps.org/doi/10.1103/PhysRevLett.95.143201}.

\bibitem[{\citenamefont{Bell et~al.}(2016)\citenamefont{Bell, Glidden, Humbert,
  Bromley, Haine, Davis, Neely, Baker, and Rubinsztein-Dunlop}}]{bell2016}
\bibinfo{author}{\bibfnamefont{T.~A.} \bibnamefont{Bell}},
  \bibinfo{author}{\bibfnamefont{J.~A.~P.} \bibnamefont{Glidden}},
  \bibinfo{author}{\bibfnamefont{L.}~\bibnamefont{Humbert}},
  \bibinfo{author}{\bibfnamefont{M.~W.~J.} \bibnamefont{Bromley}},
  \bibinfo{author}{\bibfnamefont{S.~A.} \bibnamefont{Haine}},
  \bibinfo{author}{\bibfnamefont{M.~J.} \bibnamefont{Davis}},
  \bibinfo{author}{\bibfnamefont{T.~W.} \bibnamefont{Neely}},
  \bibinfo{author}{\bibfnamefont{M.~A.} \bibnamefont{Baker}}, \bibnamefont{and}
  \bibinfo{author}{\bibfnamefont{H.}~\bibnamefont{Rubinsztein-Dunlop}},
  \bibinfo{journal}{New Journal of Physics} \textbf{\bibinfo{volume}{18}},
  \bibinfo{pages}{035003} (\bibinfo{year}{2016}),
  \urlprefix\url{https://dx.doi.org/10.1088/1367-2630/18/3/035003}.

\bibitem[{\citenamefont{Moulder et~al.}(2012)\citenamefont{Moulder, Beattie,
  Smith, Tammuz, and Hadzibabic}}]{moulder2012}
\bibinfo{author}{\bibfnamefont{S.}~\bibnamefont{Moulder}},
  \bibinfo{author}{\bibfnamefont{S.}~\bibnamefont{Beattie}},
  \bibinfo{author}{\bibfnamefont{R.~P.} \bibnamefont{Smith}},
  \bibinfo{author}{\bibfnamefont{N.}~\bibnamefont{Tammuz}}, \bibnamefont{and}
  \bibinfo{author}{\bibfnamefont{Z.}~\bibnamefont{Hadzibabic}},
  \bibinfo{journal}{Phys. Rev. A} \textbf{\bibinfo{volume}{86}},
  \bibinfo{pages}{013629} (\bibinfo{year}{2012}),
  \urlprefix\url{https://link.aps.org/doi/10.1103/PhysRevA.86.013629}.

\bibitem[{\citenamefont{Kumar et~al.}(2016)\citenamefont{Kumar, Anderson,
  Phillips, Eckel, Campbell, and Stringari}}]{kumar2016}
\bibinfo{author}{\bibfnamefont{A.}~\bibnamefont{Kumar}},
  \bibinfo{author}{\bibfnamefont{N.}~\bibnamefont{Anderson}},
  \bibinfo{author}{\bibfnamefont{W.~D.} \bibnamefont{Phillips}},
  \bibinfo{author}{\bibfnamefont{S.}~\bibnamefont{Eckel}},
  \bibinfo{author}{\bibfnamefont{G.~K.} \bibnamefont{Campbell}},
  \bibnamefont{and}
  \bibinfo{author}{\bibfnamefont{S.}~\bibnamefont{Stringari}},
  \bibinfo{journal}{New Journal of Physics} \textbf{\bibinfo{volume}{18}},
  \bibinfo{pages}{025001} (\bibinfo{year}{2016}),
  \urlprefix\url{https://dx.doi.org/10.1088/1367-2630/18/2/025001}.

\bibitem[{foo({\natexlab{a}})}]{footnote1}
\bibinfo{note}{The interaction strength $g_0=2mR\omega_\perp a_s/\hbar$, where
  $\omega_\perp$ is the frequency of the harmonic transverse confiniment and
  $a_s$ is the atomic s-wave scattering length.}

\bibitem[{foo({\natexlab{b}})}]{footnote2}
\bibinfo{note}{Note, that if $\omega/2$ is not an integer number, the original
  periodic boundary conditions become twisted boundary conditions after the
  transformation $U_p$, see \cite{SM}. For simplicity, we always choose integer
  values of $\omega/2$.}

\bibitem[{\citenamefont{Pethick and Smith}(2002)}]{Pethick2002}
\bibinfo{author}{\bibfnamefont{C.}~\bibnamefont{Pethick}} \bibnamefont{and}
  \bibinfo{author}{\bibfnamefont{H.}~\bibnamefont{Smith}},
  \emph{\bibinfo{title}{{Bose-Eistein condensation in dilute gases}}}
  (\bibinfo{publisher}{{Cambridge University Press}},
  \bibinfo{address}{{Cambridge, England}}, \bibinfo{year}{2002}).

\bibitem[{\citenamefont{Jackson and Weinstein}(2004)}]{Jackson2004}
\bibinfo{author}{\bibfnamefont{R.~K.} \bibnamefont{Jackson}} \bibnamefont{and}
  \bibinfo{author}{\bibfnamefont{M.~I.} \bibnamefont{Weinstein}},
  \bibinfo{journal}{Journal of Statistical Physics}
  \textbf{\bibinfo{volume}{116}}, \bibinfo{pages}{881} (\bibinfo{year}{2004}),
  ISSN \bibinfo{issn}{1572-9613},
  \urlprefix\url{https://doi.org/10.1023/B:JOSS.0000037238.94034.75}.

\bibitem[{\citenamefont{Mahmud et~al.}(2002)\citenamefont{Mahmud, Kutz, and
  Reinhardt}}]{Mahmud2002}
\bibinfo{author}{\bibfnamefont{K.~W.} \bibnamefont{Mahmud}},
  \bibinfo{author}{\bibfnamefont{J.~N.} \bibnamefont{Kutz}}, \bibnamefont{and}
  \bibinfo{author}{\bibfnamefont{W.~P.} \bibnamefont{Reinhardt}},
  \bibinfo{journal}{Phys. Rev. A} \textbf{\bibinfo{volume}{66}},
  \bibinfo{pages}{063607} (\bibinfo{year}{2002}),
  \urlprefix\url{https://link.aps.org/doi/10.1103/PhysRevA.66.063607}.

\bibitem[{\citenamefont{Korepin et~al.}(1993)\citenamefont{Korepin, Bogoliubov,
  and Izergin}}]{KorepinBook1993}
\bibinfo{author}{\bibfnamefont{V.~E.} \bibnamefont{Korepin}},
  \bibinfo{author}{\bibfnamefont{N.~M.} \bibnamefont{Bogoliubov}},
  \bibnamefont{and} \bibinfo{author}{\bibfnamefont{A.~G.}
  \bibnamefont{Izergin}}, \emph{\bibinfo{title}{Quantum Inverse Scattering
  Method and Correlation Functions}} (\bibinfo{publisher}{Cambridge University
  Press}, \bibinfo{address}{Cambridge, UK}, \bibinfo{year}{1993}).

\bibitem[{\citenamefont{Gaudin}(2014)}]{gaudin_2014}
\bibinfo{author}{\bibfnamefont{M.}~\bibnamefont{Gaudin}},
  \emph{\bibinfo{title}{The Bethe Wavefunction}} (\bibinfo{publisher}{Cambridge
  University Press}, \bibinfo{year}{2014}).

\bibitem[{\citenamefont{Castin}(2001)}]{Castin_LesHouches}
\bibinfo{author}{\bibfnamefont{Y.}~\bibnamefont{Castin}}, in
  \emph{\bibinfo{booktitle}{Coherent atomic matter waves}}, edited by
  \bibinfo{editor}{\bibfnamefont{R.}~\bibnamefont{Kaiser}},
  \bibinfo{editor}{\bibfnamefont{C.}~\bibnamefont{Westbrook}},
  \bibnamefont{and} \bibinfo{editor}{\bibfnamefont{F.}~\bibnamefont{David}}
  (\bibinfo{publisher}{Springer Berlin Heidelberg}, \bibinfo{address}{Berlin,
  Heidelberg}, \bibinfo{year}{2001}), pp. \bibinfo{pages}{1--136}, ISBN
  \bibinfo{isbn}{978-3-540-45338-3}.

\bibitem[{\citenamefont{Zi\'n et~al.}(2008)\citenamefont{Zi\'n, Chwede\'nczuk,
  Ole\'s, Sacha, and Trippenbach}}]{Zin2008}
\bibinfo{author}{\bibfnamefont{P.}~\bibnamefont{Zi\'n}},
  \bibinfo{author}{\bibfnamefont{J.}~\bibnamefont{Chwede\'nczuk}},
  \bibinfo{author}{\bibfnamefont{B.}~\bibnamefont{Ole\'s}},
  \bibinfo{author}{\bibfnamefont{K.}~\bibnamefont{Sacha}}, \bibnamefont{and}
  \bibinfo{author}{\bibfnamefont{M.}~\bibnamefont{Trippenbach}},
  \bibinfo{journal}{EPL (Europhysics Letters)} \textbf{\bibinfo{volume}{83}},
  \bibinfo{pages}{64007} (\bibinfo{year}{2008}),
  \urlprefix\url{http://stacks.iop.org/0295-5075/83/i=6/a=64007}.

\bibitem[{\citenamefont{Ribeiro et~al.}(2008)\citenamefont{Ribeiro, Vidal, and
  Mosseri}}]{Ribeiro2008}
\bibinfo{author}{\bibfnamefont{P.}~\bibnamefont{Ribeiro}},
  \bibinfo{author}{\bibfnamefont{J.}~\bibnamefont{Vidal}}, \bibnamefont{and}
  \bibinfo{author}{\bibfnamefont{R.}~\bibnamefont{Mosseri}},
  \bibinfo{journal}{Phys. Rev. E} \textbf{\bibinfo{volume}{78}},
  \bibinfo{pages}{021106} (\bibinfo{year}{2008}),
  \urlprefix\url{https://link.aps.org/doi/10.1103/PhysRevE.78.021106}.

\bibitem[{\citenamefont{Ole\'s et~al.}(2010)\citenamefont{Ole\'s, Zi\'n,
  Chwede\'nczuk, Sacha, and Trippenbach}}]{Oles2010}
\bibinfo{author}{\bibfnamefont{B.}~\bibnamefont{Ole\'s}},
  \bibinfo{author}{\bibfnamefont{P.}~\bibnamefont{Zi\'n}},
  \bibinfo{author}{\bibfnamefont{J.}~\bibnamefont{Chwede\'nczuk}},
  \bibinfo{author}{\bibfnamefont{K.}~\bibnamefont{Sacha}}, \bibnamefont{and}
  \bibinfo{author}{\bibfnamefont{M.}~\bibnamefont{Trippenbach}},
  \bibinfo{journal}{Laser Physics} \textbf{\bibinfo{volume}{20}},
  \bibinfo{pages}{671} (\bibinfo{year}{2010}), ISSN \bibinfo{issn}{1555-6611},
  \urlprefix\url{http://dx.doi.org/10.1134/S1054660X10050130}.

\bibitem[{\citenamefont{Dziarmaga}(2004)}]{Dziarmaga2004}
\bibinfo{author}{\bibfnamefont{J.}~\bibnamefont{Dziarmaga}},
  \bibinfo{journal}{Phys. Rev. A} \textbf{\bibinfo{volume}{70}},
  \bibinfo{pages}{063616} (\bibinfo{year}{2004}),
  \urlprefix\url{https://link.aps.org/doi/10.1103/PhysRevA.70.063616}.

\bibitem[{\citenamefont{Weiss and Castin}(2009)}]{Weiss2009}
\bibinfo{author}{\bibfnamefont{C.}~\bibnamefont{Weiss}} \bibnamefont{and}
  \bibinfo{author}{\bibfnamefont{Y.}~\bibnamefont{Castin}},
  \bibinfo{journal}{Phys. Rev. Lett.} \textbf{\bibinfo{volume}{102}},
  \bibinfo{pages}{010403} (\bibinfo{year}{2009}),
  \urlprefix\url{https://link.aps.org/doi/10.1103/PhysRevLett.102.010403}.

\bibitem[{\citenamefont{Sacha et~al.}(2009)\citenamefont{Sacha, M\"uller,
  Delande, and Zakrzewski}}]{Sacha2009}
\bibinfo{author}{\bibfnamefont{K.}~\bibnamefont{Sacha}},
  \bibinfo{author}{\bibfnamefont{C.~A.} \bibnamefont{M\"uller}},
  \bibinfo{author}{\bibfnamefont{D.}~\bibnamefont{Delande}}, \bibnamefont{and}
  \bibinfo{author}{\bibfnamefont{J.}~\bibnamefont{Zakrzewski}},
  \bibinfo{journal}{Phys. Rev. Lett.} \textbf{\bibinfo{volume}{103}},
  \bibinfo{pages}{210402} (\bibinfo{year}{2009}),
  \urlprefix\url{https://link.aps.org/doi/10.1103/PhysRevLett.103.210402}.

\bibitem[{\citenamefont{Cao et~al.}(2022)\citenamefont{Cao, Sajjad, Mas,
  Simmons, Tanlimco, Nolasco-Martinez, Shimasaki, Kondakci, Galitski, and
  Weld}}]{Cao2022}
\bibinfo{author}{\bibfnamefont{A.}~\bibnamefont{Cao}},
  \bibinfo{author}{\bibfnamefont{R.}~\bibnamefont{Sajjad}},
  \bibinfo{author}{\bibfnamefont{H.}~\bibnamefont{Mas}},
  \bibinfo{author}{\bibfnamefont{E.~Q.} \bibnamefont{Simmons}},
  \bibinfo{author}{\bibfnamefont{J.~L.} \bibnamefont{Tanlimco}},
  \bibinfo{author}{\bibfnamefont{E.}~\bibnamefont{Nolasco-Martinez}},
  \bibinfo{author}{\bibfnamefont{T.}~\bibnamefont{Shimasaki}},
  \bibinfo{author}{\bibfnamefont{H.~E.} \bibnamefont{Kondakci}},
  \bibinfo{author}{\bibfnamefont{V.}~\bibnamefont{Galitski}}, \bibnamefont{and}
  \bibinfo{author}{\bibfnamefont{D.~M.} \bibnamefont{Weld}},
  \bibinfo{journal}{Nature Physics} \textbf{\bibinfo{volume}{18}},
  \bibinfo{pages}{1302} (\bibinfo{year}{2022}), ISSN \bibinfo{issn}{1745-2481},
  \urlprefix\url{https://doi.org/10.1038/s41567-022-01724-7}.

\bibitem[{\citenamefont{See~Toh et~al.}(2022)\citenamefont{See~Toh, McCormick,
  Tang, Su, Luo, Zhang, and Gupta}}]{SeeToh2022}
\bibinfo{author}{\bibfnamefont{J.~H.} \bibnamefont{See~Toh}},
  \bibinfo{author}{\bibfnamefont{K.~C.} \bibnamefont{McCormick}},
  \bibinfo{author}{\bibfnamefont{X.}~\bibnamefont{Tang}},
  \bibinfo{author}{\bibfnamefont{Y.}~\bibnamefont{Su}},
  \bibinfo{author}{\bibfnamefont{X.-W.} \bibnamefont{Luo}},
  \bibinfo{author}{\bibfnamefont{C.}~\bibnamefont{Zhang}}, \bibnamefont{and}
  \bibinfo{author}{\bibfnamefont{S.}~\bibnamefont{Gupta}},
  \bibinfo{journal}{Nature Physics} \textbf{\bibinfo{volume}{18}},
  \bibinfo{pages}{1297} (\bibinfo{year}{2022}), ISSN \bibinfo{issn}{1745-2481},
  \urlprefix\url{https://doi.org/10.1038/s41567-022-01721-w}.

\bibitem[{\citenamefont{Blanes et~al.}(2010)\citenamefont{Blanes, Casas, Oteo,
  and Ros}}]{Blanes2010}
\bibinfo{author}{\bibfnamefont{S.}~\bibnamefont{Blanes}},
  \bibinfo{author}{\bibfnamefont{F.}~\bibnamefont{Casas}},
  \bibinfo{author}{\bibfnamefont{J.~A.} \bibnamefont{Oteo}}, \bibnamefont{and}
  \bibinfo{author}{\bibfnamefont{J.}~\bibnamefont{Ros}},
  \bibinfo{journal}{European Journal of Physics} \textbf{\bibinfo{volume}{31}},
  \bibinfo{pages}{907} (\bibinfo{year}{2010}),
  \urlprefix\url{https://doi.org/10.1088\%2F0143-0807\%2F31\%2F4\%2F020}.

\bibitem[{\citenamefont{Giergiel et~al.}(2018)\citenamefont{Giergiel, Kosior,
  Hannaford, and Sacha}}]{Giergiel2018a}
\bibinfo{author}{\bibfnamefont{K.}~\bibnamefont{Giergiel}},
  \bibinfo{author}{\bibfnamefont{A.}~\bibnamefont{Kosior}},
  \bibinfo{author}{\bibfnamefont{P.}~\bibnamefont{Hannaford}},
  \bibnamefont{and} \bibinfo{author}{\bibfnamefont{K.}~\bibnamefont{Sacha}},
  \bibinfo{journal}{Phys. Rev. A} \textbf{\bibinfo{volume}{98}},
  \bibinfo{pages}{013613} (\bibinfo{year}{2018}),
  \urlprefix\url{https://link.aps.org/doi/10.1103/PhysRevA.98.013613}.

\bibitem[{\citenamefont{Sacha}(2015{\natexlab{b}})}]{sacha15a}
\bibinfo{author}{\bibfnamefont{K.}~\bibnamefont{Sacha}}, \bibinfo{journal}{Sci.
  Rep.} \textbf{\bibinfo{volume}{5}}, \bibinfo{pages}{10787}
  (\bibinfo{year}{2015}{\natexlab{b}}),
  \urlprefix\url{https://www.nature.com/articles/srep10787}.

\bibitem[{\citenamefont{{Hannaford} and {Sacha}}(2022)}]{Hannaford2022}
\bibinfo{author}{\bibfnamefont{P.}~\bibnamefont{{Hannaford}}} \bibnamefont{and}
  \bibinfo{author}{\bibfnamefont{K.}~\bibnamefont{{Sacha}}},
  \bibinfo{journal}{Association of Asia Pacific Physical Societies Bulletin}
  \textbf{\bibinfo{volume}{32}}, \bibinfo{eid}{12} (\bibinfo{year}{2022}),
  \eprint{2202.05544}.

\bibitem[{\citenamefont{Guo et~al.}(2013)\citenamefont{Guo, Marthaler, and
  Sch\"on}}]{Guo2013}
\bibinfo{author}{\bibfnamefont{L.}~\bibnamefont{Guo}},
  \bibinfo{author}{\bibfnamefont{M.}~\bibnamefont{Marthaler}},
  \bibnamefont{and} \bibinfo{author}{\bibfnamefont{G.}~\bibnamefont{Sch\"on}},
  \bibinfo{journal}{Phys. Rev. Lett.} \textbf{\bibinfo{volume}{111}},
  \bibinfo{pages}{205303} (\bibinfo{year}{2013}),
  \urlprefix\url{https://link.aps.org/doi/10.1103/PhysRevLett.111.205303}.

\bibitem[{\citenamefont{Guo}(2021)}]{GuoBook2021}
\bibinfo{author}{\bibfnamefont{L.}~\bibnamefont{Guo}},
  \emph{\bibinfo{title}{Phase Space Crystals}}, 2053-2563
  (\bibinfo{publisher}{IOP Publishing}, \bibinfo{year}{2021}), ISBN
  \bibinfo{isbn}{978-0-7503-3563-8},
  \urlprefix\url{https://dx.doi.org/10.1088/978-0-7503-3563-8}.

\bibitem[{\citenamefont{Giergiel et~al.}(2022)\citenamefont{Giergiel, Lier,
  Sur\'owka, and Kosior}}]{Giergiel2022}
\bibinfo{author}{\bibfnamefont{K.}~\bibnamefont{Giergiel}},
  \bibinfo{author}{\bibfnamefont{R.}~\bibnamefont{Lier}},
  \bibinfo{author}{\bibfnamefont{P.}~\bibnamefont{Sur\'owka}},
  \bibnamefont{and} \bibinfo{author}{\bibfnamefont{A.}~\bibnamefont{Kosior}},
  \bibinfo{journal}{Phys. Rev. Res.} \textbf{\bibinfo{volume}{4}},
  \bibinfo{pages}{023151} (\bibinfo{year}{2022}),
  \urlprefix\url{https://link.aps.org/doi/10.1103/PhysRevResearch.4.023151}.

\end{thebibliography}
\end{document}